%% file: main.tex
\documentclass[sigconf]{acmart}

\usepackage{enumitem}
\usepackage{makecell}
\usepackage{multirow}
\usepackage{multicol}
\usepackage{bbding}
\usepackage{pifont} 
\usepackage{colortbl}
\usepackage{booktabs}
\usepackage[edges]{forest}

\begin{document}

\title{Multimodal Financial Foundation Models (MFFMs): \\ Progress, Prospects, and Challenges}


\author{Xiao-Yang Liu Yanglet}
\affiliation{%
  \institution{ Columbia University}
  \city{New York}
  \state{NY}
  \country{USA}}
\email{XL2427@columbia.edu}

\author{Yupeng Cao}
\affiliation{%
  \institution{Stevens Institute of Technology}
  \city{Hoboken}
  \state{NJ}
  \country{USA}}
\email{caoyupeng.work@gmail.com}

\author{Li Deng}
\affiliation{
  \institution{University of Washington}
  \city{Seattle}
  \state{WA}
  \country{USA}
}
\email{deng629@gmail.com}

\setcopyright{acmlicensed}

\copyrightyear{2024}
\acmYear{2024}
\setcopyright{acmlicensed}
\acmConference[ICAIF 24]{International Workshop on Multimodal Financial Foundation Models (MFFMs) at 5th ACM International Conference on AI in Finance}{Nov. 15}{NY, USA}
\acmBooktitle{International Workshop on Multimodal Financial Foundation Models (MFFMs) at ACM International Conference on AI in Finance (MFFM at ICAIF), Nov. 14--17, 2024,}

\begin{abstract}

Financial Large Language Models (FinLLMs), such as open FinGPT and proprietary BloombergGPT, have demonstrated great potential in select areas of financial services. Beyond this earlier language-centric approach, Multimodal Financial Foundation Models (MFFMs) can digest interleaved multimodal financial data, including fundamental data, market data, data analytics, macroeconomic, and alternative data (e.g., natural language, audio, images, and video). In this position paper, presented at the MFFM Workshop joined with the ACM International Conference on AI in Finance (ICAIF) 2024, we describe the progress, prospects, and challenges of MFFMs. This paper also highlights ongoing research in the \textbf{SecureFinAI Lab}\footnote{\url{https://openfin.engineering.columbia.edu/}} at Columbia University and summarizes the FinLLM Exploration meetings at FinOS of Linux Foundation. MFFMs will enable users to better understand the underlying complexity associated with numerous financial tasks and data, simplifying the operation of financial services and investment processes. \\
\textcolor{blue}{\textbf{Github: \url{https://github.com/Open-Finance-Lab/Awesome-MFFMs/}}}

\end{abstract}

\begin{teaserfigure}
\centering
  \includegraphics[width=0.99\textwidth, trim={0.0cm 0.70cm 0cm 0.1cm}, clip]{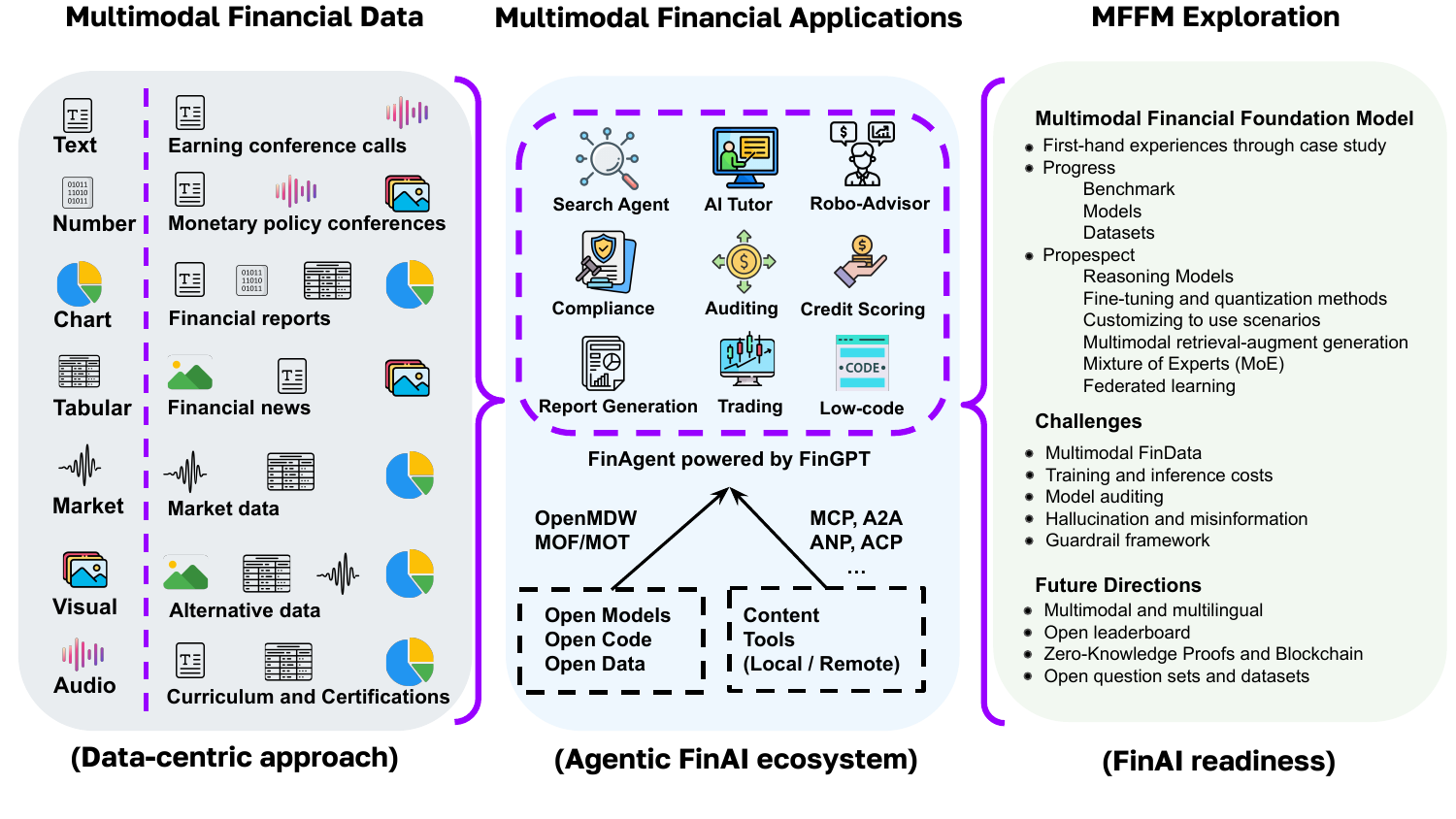}
  \caption{An overview. Multimodal financial data (left block) is ubiquitous in real life, while financial applications (middle block) require financial agents to possess multimodal capabilities, such as search, tutor, robo-advisor, compliance, auditing, trading, and low-code development. However, the urgent need for MFFM models (right block) call for immediate actions, in order to achieve FinAI readiness and governance.
  }
  \label{fig:overall}
  \vspace{-0.05in}
\end{teaserfigure}
\maketitle

\section{Introduction}

\textit{The general public could not afford a private lunch with billionaire Warren Buffett. What about hiring Buffett as my on-call financial advisor, possibly at a cost of \$100?  Before finalizing an investment of \$50 million, how about holding an elite brainstorming session with the world’s 10 greatest investors?...at a cost of \$5000?... Not many years later, as he earned his first \$1 million from investment, Yanglet was to remember that night when Dr. Li Deng helped him revise the proposal for the Multimodal Financial Foundation Models (MFFMs) workshop.} 
\\

Generative AI is witnessing the rise of foundation models (e.g., transformer neural network, diffusion models) trained on massive data that can be adapted to a wide range of downstream tasks \cite{bommasani2021opportunities}. Recently, large language models (LLMs) have demonstrated remarkable proficiency in understanding and generating human-like texts. FinLLMs, such as open FinGPT \cite{Liu2023FinGPT, Liu2024FinGPTHPC, Yang2023FinGPT,zhang2023instruct, Zhang2023RetrievalAugmentedLLMs, Felix2024FinGPTAgent} and proprietary BloombergGPT \cite{Wu2023BloombergGPT}, have shown great potential in select areas of financial services \cite{kong2024large, nie2024survey}. Beyond this earlier language-centric approach, Multimodal Financial Foundation Models (MFFMs) can process interleaved multimodal financial data, including fundamental data, market data, data analytics, macroeconomics, and alternative data (e.g. natural language, audio, visual). Multimodal financial data has unique characteristics, such as dynamic, both structured and unstructured forms, and comes in varying formats (e.g., charts, graphs, Web APIs, Excel spreadsheets, SEC filings \cite{islam2023financebench}, XBRL filings \cite{Han2024XBRLAgent}, and SQL data \cite{zhao2024revolutionizing}). 


On the journey toward widespread adoption of MFFMs, several challenges remain, including increasing concerns related to reproducibility, transparency, privacy, and ethics. First, many existing LLMs function as black boxes, posing challenges in comprehending their operations and ensuring fairness. Another two major challenges are ``model cannibalism" and ``openwashing." Many models are largely trained and released without transparency in mind, e.g., the Claude 3.5 Sonnet. Many supposedly "novel" models may exploit labels from existing LLMs (e.g., GPT-4o) and perform supervised learning, referred to as ``model cannibalism." As a result, MFFMs are opaque in decision-making. They give rise to various challenges, such as inadequate transparency in training data, deficiencies in combating models’ inherent biases, safety and security issues, and adversarial attacks (for example, backdoor attacks).

Recently, there have been many ``openwashing" behaviors, where the LLM weights are marketed as "open" but under restricted licenses, while OSI-approved licenses (e.g., Apache License 2.0 and MIT License) are preferable. A research alliance between Columbia University, Oxford University, and the Linux Foundation proposed the Model Openness Framework \cite{white2024model} that classifies model openness by ranking models according to $16$ components. This framework offers guidance to researchers and model producers for promoting transparency and reproducibility. MFFMs that comply with this framework would promote reproducibility and adoption across the finance industry, such as Open FinLLMs \cite{Xie2024OpenFinLLMs}. For financial institutions, it provides clear guidelines for new models to become commercially suitable without restrictions.

In this position paper, presented at the MFFM Workshop\footnote{MFFM Workshop: \url{https://sites.google.com/view/iwmffm2024/home?authuser=1}} jointly held with ACM International Conference on AI in Finance (ICAIF) 2024, we describe the progress, prospects, and challenges of MFFMs. This paper also highlights ongoing research in the SecureFinAI Lab at Columbia University and and summarizes the FinLLM Exploration meetings at FinOS of Linux Foundation. We first list multimodal financial data and data-centric approach in the financial domain (left block of Fig. \ref{fig:overall}). Then, we describe multimodal financial applications (middle block of Fig. \ref{fig:overall}). We envision that FinAgent is a promising solution to build multiple financial applications. Following the Model Openness Framework, open licenses (e.g. OpenMDW\footnote{https://openmdw.ai/}) and agent protocols (e.g. Model Context Protocol, Agent2Agent Protocol) enables the blooming of an agentic FinAI ecosystem. However, several major challenges (right block of Fig. \ref{fig:overall}) call for immediate actions in order to achieve FinAI readiness. The associated challenges are proprietary data constraints, training and inference costs, regulatory complexities, reasoning capacity, and the need for robust benchmarks and a guardrail framework to address misinformation and data biases. We believe that MFFMs will enable promising financial tasks and data analysis, streamlining the operation of financial services.

\begin{table*}[t]
\centering
\begin{tabular}{llcccccc}
\toprule
Survey  & Date & FinLLMs & Benchmark & Applications & Challenges & Multimodal & Readiness/Governance      \\ \midrule
\citet{li2023large} 
&  Nov. 2023 & \textcolor{green}{\CheckmarkBold}     & \textcolor{red}{\XSolidBrush}        &  \ding{111}      & \ding{111}        & \textcolor{red}{\XSolidBrush}         & \textcolor{red}{\XSolidBrush}    \\

\citet{ding2024large}
&  Jul. 2024 & \textcolor{red}{\XSolidBrush}     & \textcolor{red}{\XSolidBrush} & \ding{111} & \ding{111} & \textcolor{red}{\XSolidBrush} & \textcolor{red}{\XSolidBrush}    
\\
\citet{lee2024survey}  &  Apr. 2024 & \textcolor{green}{\CheckmarkBold}     & \textcolor{green}{\CheckmarkBold}     & \ding{111}        & \ding{111}        & \textcolor{red}{\XSolidBrush}       & \textcolor{red}{\XSolidBrush}      \\
\citet{kong2024large, nie2024survey}  &  Late 2024   & \textcolor{green}{\CheckmarkBold}     & \textcolor{green}{\CheckmarkBold}       & \textcolor{green}{\CheckmarkBold}         & \ding{111}       & \ding{111}       & \textcolor{red}{\XSolidBrush}         \\ \midrule
\textbf{This Survey} &  Jan. 2025 & \textcolor{green}{\CheckmarkBold}     & \textcolor{green}{\CheckmarkBold}       & \textcolor{green}{\CheckmarkBold}        & \textcolor{green}{\CheckmarkBold}       & \textcolor{green}{\CheckmarkBold}      & \textcolor{green}{\CheckmarkBold}    \\ \bottomrule
\end{tabular}
\caption{Overview of related surveys. The square indicates that the topic was covered but not comprehensive.}
\label{tab:survey}
\vspace{-1em}
\end{table*}

\begin{table*}[t]
\centering
\begin{tabular}{l|c|c|c|c|c|c|c|c}
\toprule
\multicolumn{1}{c|}{\textbf{Types}} & \textbf{Text} & \textbf{Audio} & \textbf{Image} & \textbf{Video} & \textbf{Numbers} & \textbf{Tabular} & \textbf{Chart} & \textbf{Time-Series} \\ \midrule
Earnings Conference Calls (ECC)                              
& \textcolor{green}{\CheckmarkBold} & \textcolor{green}{\CheckmarkBold} &  &  &  &  & &                  \\
Monetary Policy Calls (MPC)                                  
& \textcolor{green}{\CheckmarkBold} & \textcolor{green}{\CheckmarkBold} &  & \textcolor{green}{\CheckmarkBold} &  & &  &  \\
Climate Data                                           
& \textcolor{green}{\CheckmarkBold} & & \textcolor{green}{\CheckmarkBold}  &   &  &  &  &                  \\ 
Financial News & \textcolor{green}{\CheckmarkBold} &                & \textcolor{green}{\CheckmarkBold}  &                &     \textcolor{green}{\CheckmarkBold}           &                  &  &                  \\
Market Data                                     
& \textcolor{green}{\CheckmarkBold} & & &  & \textcolor{green}{\CheckmarkBold} & \textcolor{green}{\CheckmarkBold} & \textcolor{green}{\CheckmarkBold} & \textcolor{green}{\CheckmarkBold}  \\
Financial Reports
& \textcolor{green}{\CheckmarkBold} & &  \textcolor{green}{\CheckmarkBold} &  & \textcolor{green}{\CheckmarkBold} & \textcolor{green}{\CheckmarkBold} & \textcolor{green}{\CheckmarkBold} &  \textcolor{green}{\CheckmarkBold} \\
Financial curriculum and certifications                              
& \textcolor{green}{\CheckmarkBold} &  &  &  & \textcolor{green}{\CheckmarkBold} & \textcolor{green}{\CheckmarkBold} & \textcolor{green}{\CheckmarkBold} &                  \\
\bottomrule
\end{tabular}
\caption{Overview of multimodal financial data.}
\label{tab:multimodal_data}
\vspace{-0.1in}
\end{table*}

\textbf{Related Work}: Several related surveys are given in Table~\ref{tab:survey}. \citet{li2023large} first reviewed the approach employing LLMs in finance. \citet{ding2024large} summarized the performance of LLM-based agents in financial trading tasks. \citet{lee2024survey} reviewed FinLLMs from a benchmark perspective. \citet{kong2024large} and \citet{nie2024survey} further summarize recent advancements in FinLLMs and discuss their various application scenarios.
Despite these efforts, the rapidly evolving nature of the field necessitates an updated and thorough review of multimodal financial data, applications, and models. Furthermore, there is still a need for an in-depth analysis concerning the opportunities and challenges of applying LLMs in finance, including their current readiness level.

\textbf{Contributions}: We summarize the contributions as follows: 
\vspace{-0.5em}
\begin{itemize}[leftmargin=*]
    \item To the best of our knowledge, this is the first comprehensive survey of MFFMs. We summarize three aspects, multimodal financial data, applications, and model exploration.
    \item For multimodal financial data, we emphasize a data-centric approach. For applications, we point out the promising era of an agentic AI ecosystem with various types of FinAgents, which is enabled by open models and agent protocols.  
    \item We compare and contrast MFFMs with LLMs, FinLLMs, and MM-LLMs. Our aim is to offer readers a holistic view of the MFFM development lifecycle and help readers understand current progress and future prospects.
    \item We describe the opportunities and point out the challenges when applying MFFMs in the financial domain, including proprietary data and digital regulatory reporting.
    \item We discuss ethical challenges of MFFMs's readiness, including the hallucination and misinformation in use scenarios, as well as the importance of building a guardrail framework.
\end{itemize}

\vspace{0.1in}
\noindent \textit{Well, using an MFFM base model, one could fine-tune a ``Warren Buffett" agent\footnote{Demo of Buffet agent: \url{https://finlora-docs.readthedocs.io/en/latest/intro/demo.html}} using the FinLoRA method \cite{wang2024finlora} as well as QLoRA method \cite{dettmers2023qlora} by feeding multimodal data \cite{Liu2023FinGPT, Liu2024FinGPTHPC}, e.g., Buffett‘s conference transcripts, audio, video, interviews, and the fine-tuning cost would be less than \$100. On the other hand, by specifying the preferred articles/websites or creating a customized database as in \cite{Felix2024FinGPTAgent}, an investment institution would consult the world’s ten greatest investors (namely their digital avatars) in an elite brainstorming session at the cost of \$5000.}

\section{Data-Centric Approach for Multimodal Financial Data}

We first summarize the common multimodal financial data in Section~\ref{ssec:spectrum_mmdata}. Then, we describe typical types of multimodal financial data in Sections 2.2 to Section 2.7.

\subsection{Spectrum of Multimodal Financial Data}
\label{ssec:spectrum_mmdata}

 We emphasize the data-centric approach \cite{mazumder2023dataperf, zha2025data}, since \textbf{data readiness is a prerequisite of AI readiness}. The principle in computing, ``garbage in, garbage out" (also known as GIGO), states that flawed, biased, or poor quality ("garbage") data or input produces a result or output of similar ("garbage") quality. If feed a model with inaccurate, incomplete, or irrelevant data, one can expect the results it produces to be equally flawed. Therefore, one cannot rely on such a model for decision-making processes in finance. 

 Multimodal data are common in business, finance, accounting, and auditing, as illustrated in Fig.~\ref{fig:overall} (left block).
\begin{itemize}[leftmargin=*]
    \item \textbf{Text data}: Text is the most prevalent data type, including financial news, financial reports, transcripts of earnings conference call, and social media posts. These textual data provide timely market information and reflect the sentiments of market participants.
    
    \item \textbf{Number}: Numerical data, such as stock prices, financial indicators, and economic statistics, offer market insights. Analysts and investors frequently rely on numerical data for market forecasting.
    
    \item \textbf{Chart data}: Charts are frequently included in financial reports, news articles, and related materials. It visually represents market trends and patterns, facilitating easier 
    interpretation of market behavior and dynamics.
    
    \item \textbf{Tabular data:} Structured financial data presented in tables, including balance sheets, income statements, stock prices, and trading volumes. 
    
    \item \textbf{Market data}: It is a sequence of data points indexed in time order. In the financial sector, time series data is commonly used to represent how a financial indicator changes over time.
    
    \item \textbf{Visual data}: Visual data includes images and videos. They are from financial media and official announcements. Visual data provides detailed insights beyond textual and numerical data, illustrating complex market events and trends.
    
    \item \textbf{Audio data:} Financial podcasts and recordings of earnings conference calls contain critical auditory information. Audio modalities can influence market perception and offer additional dimensions for sentiment analysis and market prediction.

\end{itemize}

Multimodal financial data can refer to a combination of the above uni-modal data. For instance, Earning Conference Calls (ECCs) consist of two modalities: the audio of a presentation and its textual transcripts.  Multimodal financial data has several unique characteristics, including data streaming, low signal-to-noise ratio (Low-SNR), and semistructured formats \cite{de2018advances}. We list the common types in Table~\ref{tab:multimodal_data} and describe them in the following subsections.

\subsection{Earning Conference Calls (ECCs)}

\begin{figure}[t]
\begin{center}
\centerline{\includegraphics[width=\columnwidth]{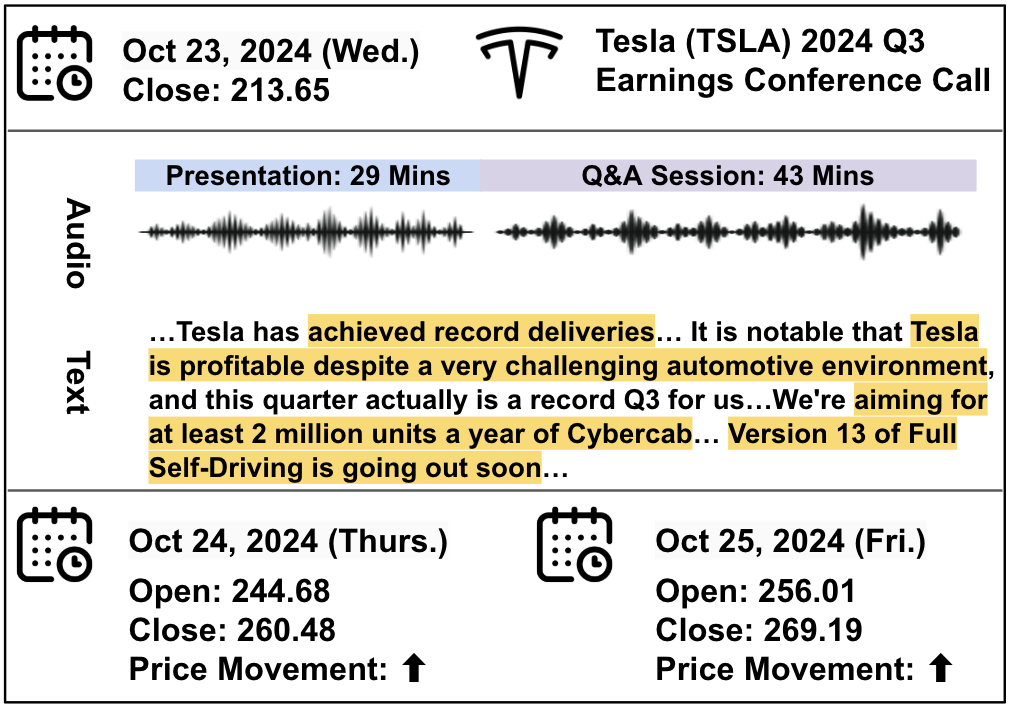}}
\caption{An ECC example of Tesla 2024 Q3 on Oct. 23, 2024. The CEO, Elon Musk, presented a speech to explain the company's revenue for the past quarter and major related events and provided an overview plan. The close prices of the following two days were $\$260.48$ and $\$ 269.19$.}
\label{fig:ecc}
\end{center}
\vspace{-2em}
\end{figure}

The earnings conference call (ECC) is a teleconference or webcast held quarterly by a public company. Stakeholders (including analysts, investors, and the media) participate to obtain the latest financial status of the company. First, the CEO/CFO highlights the quarterly financial status, strategic initiatives, and forward-looking plans. Then, analysts and investors ask questions during the Q\&A session. The release of ECCs is correlated with market reactions, making them an important resource for analyzing market changes~\cite{foster1984earnings}.

An ECC example of Tesla 2024 Q3 is shown in Fig. ~\ref{fig:ecc}. This call has 72 minutes, including a 29 minute presentation by Tesla CEO Elon Musk and 43 minute Q\&A session. First, Elon Musk summarized Tesla's Q3 revenue and car production status and underscored Tesla's ongoing strategy to accelerate the global transition to sustainable energy. In the end, Elon reiterated Tesla's preparations for introducing more affordable models. During the Q\&A session, Tesla's executive team responded to questions about product research and development, upcoming product plans, Tesla's Full Self-Driving offerings, etc. Owing to good revenue performance and car production, Tesla's stock price sustained an upward trend in the following two days. The entire ECC is saved as an ".mp3/.wav" audio file, and the corresponding transcript is also recorded. Both audio and text data can be accessed and analyzed by the general public.

The creation of an ECC dataset is critical for developing analytical tools, particularly for stock movement prediction and risk modeling. MDRM \cite{qin2019you} is a representative ECC dataset that includes 576 earnings conference calls from 280 companies in the S\&P 500 for the year 2017. The entire dataset is 5.7 GB in storage. The author segmented the transcripts into individual sentences and aligned them with corresponding audio clips, resulting in a total of 88,829 paired sentences and audio clips. The audio data is available from \textsc{earningcast}\footnote{https://earningscast.com/}
and transcript file can be downloaded from \textsc{Seeking Alpha}\footnote{https://seekingalpha.com/}. A series of research works extract critical information from textual transcripts and integrate it with audio features such as tone and sentiment in a speech to assess risks (e.g., volatility)~\cite{qin2019you, yang2020html, cao2024ecc}.

The current approach to financial analysis based on ECC data faces several key challenges related to dataset curation. First, the existing ECC dataset is limited in size and lacks sufficient coverage of companies across diverse industries, as ECC characteristics vary considerably between companies. Second, aligning audio with the text remains imperfect. Splitting the ECC data into segments frequently fails to align precisely with sentence boundaries, potentially leading to semantic incoherence in the segmented text and audio. Therefore, it is essential to establish a dataset curation pipeline that focuses on acquiring, organizing, segmenting, and labeling ECC data. Such a data infrastructure will enable the creation of more effective financial applications.

\begin{table*}[htbp]
\centering
\resizebox{\textwidth}{!}{
\begin{tabular}{|l|l|l|l|c|}
\hline
\textbf{Financial Reports} & \textbf{Frequency} & \textbf{Publisher} & \textbf{Focus}& \textbf{Required by SEC}  \\
\hline
Form 10-Q & Quarterly & Company & Interim financial statements and recent operational updates & Yes \\
\hline
Form 10-K & Annually &  Company &  Comprehensive, audited financial results and business overview & Yes \\

\hline
DEF 14A & Annually & Company & Governance, executive compensation, and voting matters & Yes \\
\hline
Form 8-K & Event-driven & Company & Disclosure of significant, material events affecting operations & Yes \\
\hline
Earnings Release & Quarterly & Company & Preliminary quarterly financial results and management commentary & No \\
\hline
Annual Report & Annually & Company & Simplified summary highlighting performance and management vision & No  \\
\hline
Zacks Investment Reports & Frequently & Third-party (Zacks) & Stock ratings, earnings forecasts, and investment recommendations & No  \\
\hline
Sell-side Broker Reports & Frequently & Third-party (Analysts) & In-depth analysis, valuation, forecasts, and buy/sell recommendations & No \\
\hline
\end{tabular}}
\caption{Financial reports with different frequency, source, primary focus and SEC requirements.}
\label{tab:financial_reports}
\vspace{-1em}
\end{table*}

\subsection{Monetary Policy Conferences (MPCs)}
Monetary Policy Conferences (MPCs) are regularly held by a country’s central bank, like the U.S. Federal Reserve. An MPC deliberates on a nation's economic conditions, articulates monetary policy, and assesses potential economic risks. The conference includes a press presentation by the governor, followed by a Q\&A session with journalists. Given their provision of critical insights into a central bank's decisions, MPCs are instrumental in influencing economic conditions, such as commodity trading, inflation, and currency exchange rates.

\begin{figure}[t]
\begin{center}
\centerline{\includegraphics[width=\columnwidth]{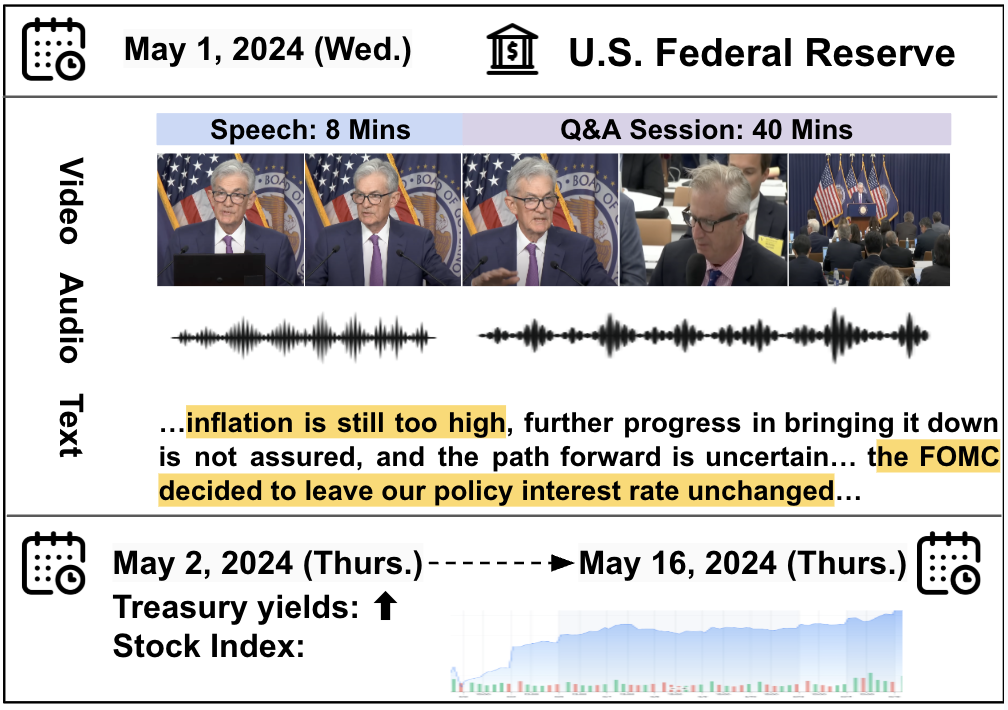}}
\caption{An MPC example held by the U.S. Federal Reserve on May 1st, 2024. The Governor presented a press speech, followed by a Q\&A session. After the conference, the stock index declined, while Treasury yields increased. Later, the stock index exhibited a gradual recovery.}
\label{fig:mpc}
\end{center}
\vskip -0.2in
\end{figure}

Fig. ~\ref{fig:mpc} shows an example of the Federal Open Market Committee (FOMC) conference on May 1st, 2024. This conference has 48 minutes with 8 minutes presentation and 40 minutes Q\&A session. In the presentation, Federal Reserve Chairman Jerome Powell emphasized that inflation remains a major concern and suggested the possibility of further policy tightening. Then, the FOMC unanimously decided to keep the benchmark short-term borrowing rate at 5.25\%–5.5\%. This rate is the highest level in 23 years. After the release of the conference recording, consumer surveys indicated heightened economic anxiety\footnote{Reported by CNBC: https://www.cnbc.com/2024/05/22/fed-minutes-may-2024-.html}. Stock markets experienced declines while Treasury yields increased. In the subsequent period, however, the stock index gradually recovered.

Analyzing MPC data can help illuminate policy decisions and assist in forecasting economic trends. The first comprehensive MPC dataset MONOPOLY~\cite{mathur2022monopoly} has 180 GB in size. It includes 340 MPC instances from six countries' central banks: the United States, the United Kingdom, the European Union, Canada, New Zealand, and South Africa. In total, the dataset comprises 15,729 minutes of recorded content, where each MPC session has on average 53 minutes. Typically, every MPC includes approximately 10 minutes of presentation, followed by a Q\&A session with more than 40 minutes. Each MPC consists of three parts: Audio, Text, and Video.  The dataset was constructed by employing the BeautifulSoup Python package\footnote{https://www.crummy.com/software/BeautifulSoup/} to scrape MPC dates, `.mp3' audio, `.MP4' videos, and PDF transcripts. Text data are subsequently extracted from PDF transcripts by using Urllib\footnote{https://pypi.org/project/urllib3/}. Recent studies use this dataset to jointly model audio, text, and video features to predict various economic indicators~\cite{mathur2022monopoly,ouyang2024modal}.

The following challenges are faced when using the MPC dataset to build an analysis tool for economic conditions. First, storing MPC data needs to maintain audio, text, and video modalities together. It introduces the challenges related to efficient data management. Second, processing MPC data requires precise alignment across audio, text, and video modalities, which remains technically challenging. Thus, establishing a data curation pipeline for MPC is essential for advancing related methodological development.

\subsection{Financial Reports}
The financial report is a formal document that presents a company’s financial activities, performance, management discussion, and audited financial statements. The frequently-used financial reports include filings (e.g., 10-K, 10-Q, DEF-14A, 8-K) required by the U.S. Securities and Exchange Commission (SEC), company-issued documents for stakeholders (e.g., earnings releases and annual reports), and third-party analysis reports such as Zacks reports and sell-side broker reports. These reports differ in their publication frequency, publisher, and areas of emphasis. 

A summary of these financial reports is provided in Table~\ref{tab:financial_reports}. Market participants can access various financial reports from different companies based on their specific needs. These reports enable investors to evaluate a company's status and identify broader market trends.  Additionally, these financial reports are monitored by government and regulatory agencies to ensure fairness in trading and other financial activities. Recently, on the FinanceBench dataset \cite{islam2023financebench}, GPT-4 model incorrectly answered or refused to answer 81\%
of questions.


\subsection{Financial News}

Financial news refers to news that pertains to money and investments, including news on markets. It is disseminated through various channels, including traditional financial reporting (e.g., The Wall Street Journal), financial news services platforms (e.g., Bloomberg terminal), social media (e.g., Twitter and LinkedIn), online discussion forums (e.g., Reddit), and interactive media formats such as live broadcasts. Financial news can take different formats, including text, video, audio, numerical, charts, and tabular data. It has become increasingly important in forecasting financial outcomes, such as stock volatility, investor sentiment, market risks, and macroeconomic stability~\cite{schumaker2012evaluating, pivskorec2014cohesiveness}.

The GameStop (GME) short squeeze event in January 2021 exemplifies how financial news can impact the financial markets. In the beginning, hedge funds published short-selling reports on GME, forecasting a decline in its stock price based on weak financial fundamentals. They also spread their view online, prompting institutional investors to establish short positions on GME. However, someone discovered that financial institutions were excessively short-selling GME and shared this finding on social platforms such as Reddit's r/WallStreetBets. It attracted extensive discussions, which resulted in collective buying activities. Individual investors' behavior drives GME stock prices upward. Then, Elon Musk's retweet `Gamestonk!!', making such a short squeeze event spread globally. Institutional investors subsequently faced pressure to purchase shares to cover their short positions, thereby intensifying the stock's upward momentum. The price of GME stock reached \$530 from \$1 within that two-month period, resulting in bankruptcy for a hedge fund. This incident highlights the significant impact of financial news on market trends. 

During financial events such as those previously described, large volumes of news data are generated rapidly. Efficiently managing and processing this extensive news data is a challenging task. Developing automated methods to extract useful information from massive amounts of multimodal news will save financial market practitioners a lot of time and effort. \citet{lin2024analyzing} employed LLMs to analyze GME short squeeze event. LLMs were used to clean extensive collections of online financial news, resulting in the creation of a high-quality news dataset. The dataset was then used to analyze user behavior and the underlying mechanisms driving information dissemination. Their findings underscore the substantial potential of LLMs in financial news analysis.

Effective collection of financial news data is crucial for analyzing market dynamics. Financial news data can be collected from various online platforms and sources. First, specialized financial platforms, such as Bloomberg, Dow Jones, Yahoo Finance, and CNBC, deliver timely and professional financial news. Second, professional news organizations, such as Reuters, offer financial news coverage. Third, social media, including X (formerly Twitter) and Reddit, also provide financial news. Users can access financial news data through platform-specific APIs or manually gather financial news data directly from these platforms. However, users must be mindful of copyright restrictions associated with each platform.

Although amounts of financial news data are available, there remain several challenges to making these raw data into usable datasets: 1) Trustworthiness. Financial news from various sources may include subjective content or misinformation. Evaluating the reliability of financial news is a significant challenge; 2) Volume issue. A large amount of financial news is disseminated daily, making it difficult to effectively process and manage; and 3) Modality alignment. Financial news includes various types of information, such as charts, tables, and images. A key challenge is accurately aligning textual content with its corresponding other elements.

\subsection{Market Data and Alternative Data}
\subsubsection{Market Data}
Market data refers to price information and other related data for financial instruments provided by trading venues. It represents financial information through different modalities. For example, market data uses time-series data to record a company's stock price, numerical data to represent financial indicators, and charts or tabular data to display a company’s operational performance. These multimodal market data provide investors with diverse perspectives on current market changes and historical movements, supporting informed decision-making~\cite{lee2020multimodal}. 

Financial markets have undergone a rapid change due to the increasing amount of data. Extracting actionable insights from these vast and heterogeneous market data to support decision-making in complex market environments has therefore become a challenge~\cite{hambly2023recent}. Reinforcement learning (RL) provides a promising approach to address this challenge. RL could train trading agents to interact with dynamic market environments and to optimize their financial decisions autonomously~\cite{hambly2023recent, sun2023reinforcement}. The financial reinforcement learning (FinRL) project \cite{liu2018practical, liu2020finrl, liu2021finrl, liu2022finrl} offers a user-friendly virtual market environment that includes a wide range of multimodal market data. FinRL integrates commonly used Deep Reinforcement Learning (DRL)  algorithms, enabling users to develop their own DRL trading strategies. Recently, FinRL 2025 contest\footnote{\url{https://finrl-contest.readthedocs.io/en/latest/}} proposed the FinRL-DeepSeek project, combining reinforcement learning with LLMs to develop an automated stock trading agent trained on stock price and financial news data. This hybrid approach enhances the capacity to process complex, evolving market information~\cite{wang2023alpha}.

Quantamental investment refers to combining computer-driven and human-driven research to analyze the amount of market data to construct a portfolio~\cite{tadoori2019introduction}. For example, alpha factor mining has garnered attention for its ability to identify and exploit market inefficiencies and for its seamless integration with AI-based forecasting methods.


\subsubsection{Climate Data for Commodity Trading}
Climate data is the records of climate conditions observed at specific locations and times, collected using particular instruments and standardized procedures. Common types of climate data include precipitation, temperature, wind speed, humidity, and satellite imagery of cloud coverage. Climate changes can affect the supply of goods, potentially causing significant price fluctuations and market uncertainty~\cite{stern2008economics,monasterolo2020climate}. By analyzing weather data, investors can better understand and anticipate its impact on financial markets.


\subsection{Financial Curriculum and Certifications}

\begin{figure}[t]
\begin{center}
\centerline{\includegraphics[width=\columnwidth]{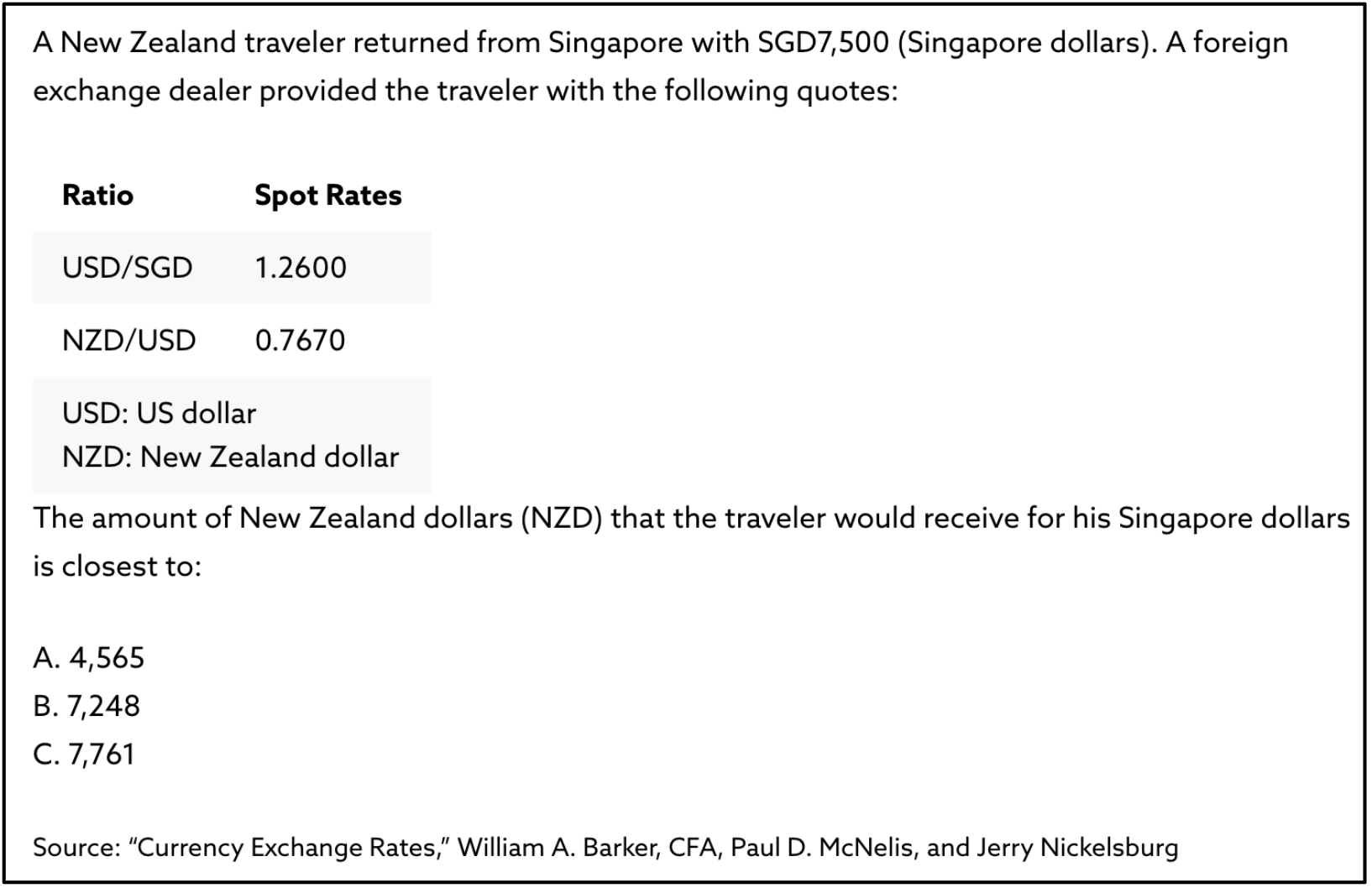}}
\caption{An example question from CFA Level-1 exam. The question is about currency exchange, where numerical data and tabular data are given. Respondents are required to comprehend the provided information, interpret the table, perform calculations, and then select the correct answer.}
\label{fig:cfa}
\end{center}
\vskip -0.2in
\end{figure}

Completion of a degree requires successfully navigating the learning path through the financial curriculum on campus. Afterward, earning a professional certification will enable a candidate to embark on her career path toward becoming a senior professional. These degree-type and certification-type questions contain multimodal financial data, including textual descriptions, numerical calculations, graphs, charts, and data tables. Correctly answering these questions requires a combination of financial knowledge and reasoning capacity. Fig. \ref{fig:cfa} shows an example of the CFA exam \footnote{\url{https://www.cfainstitute.org/programs/cfa-program/cfa-program-level-i-sample-questions}}.

Evaluating MFFMs' performance on these questions can assess whether these models truly understand financial knowledge and master the reasoning capability ~\cite{Jaisal2024QFinBen}. Recent studies have shown that ChatGPT and GPT-4 models struggle with CFA exam questions \cite{callanan2024can, mahfouz2024state}. QFinBen \cite{Jaisal2024QFinBen} and tutor agent\footnote{\url{https://finllm-leaderboard.readthedocs.io/en/latest/demos_of_finagents/tutor_agent.html}} organized university-level and professional certification problems to measure model proficiency. Moreover, such organized questions allow interpretation of the model's score. In other words, one can easily understand a model's strength and also identify where the model is struggling.

\section{Multimodal Financial Applications: \\ Agentic FinAI Ecosystem}

In this section, we describe FinAgents in real-life scenarios, which are categorized into two types, tool agents and financial service agents, respectively. We point out key enablers for an upcoming era of agentic FinAI ecosystem.

\subsection{FinAgents Powered by FinGPT}
\label{ssec:finagent}

Our SecureFinAI Lab at Columbia has developed several prototypes of FinAgents, powered by FinGPT \cite{Liu2023FinGPT, Liu2024FinGPTHPC, Yang2023FinGPT,zhang2023instruct, Felix2024FinGPTAgent}: search agent \cite{Felix2024FinGPTAgent}, tutor agent \cite{Jaisal2024QFinBen}, XBRL agent \cite{Han2024XBRLAgent}, and FinRL trading agents \cite{liu2018practical, liu2020finrl, liu2021finrl, liu2022finrl}. The search agent can retrieve real-time financial data from the Internet and generate personalized advice. The tutor agent democratizes financial knowledge and interprets complex regulations. The XBRL agent \cite{Han2024XBRLAgent} analyzes SEC filings (following the eXtensible Business Reporting Language (XBRL)) by calling an external retriever and a calculator. The FinRL trading agent \cite{liu2018practical, liu2020finrl, liu2021finrl, liu2022finrl} provides an end-to-end framework that integrates commonly used Deep Reinforcement Learning (DRL) algorithms (such as DQN, DDPG, PPO, SAC, A2C, and TD). It enables company clients to develop their trading strategies.

AI agents will enable learning systems to take action by observing the complex environment through iterative improvement~\cite{durante2024agent}. This capability could assist in addressing more complex real-world financial tasks. OpenAI\footnote{https://cdn.openai.com/business-guides-and-resources/a-practical-guide-to-building-agents.pdf} and Google~\cite{wiesinger2024agents} recently released detailed guides for agent development, which provide a good starting point for developing FinAgents.

\begin{figure*}[t]
\begin{center}
\centerline{\includegraphics[width=0.99\textwidth,  trim={0cm 0cm 0cm 0cm}, clip]{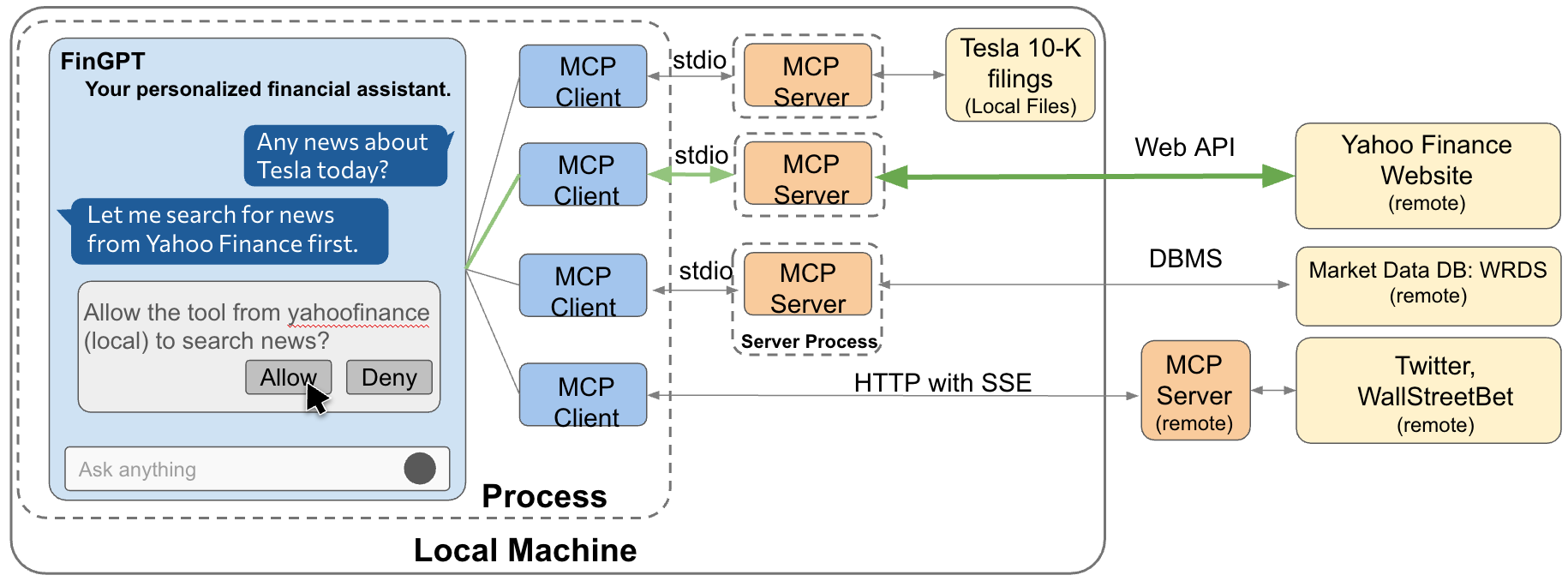}}
\caption{The FinAgent framework powered by FinGPT and agent protocols (e.g, Model Context Protocol).}
\label{fig:mcp}
\end{center}
\end{figure*}


Fig. \ref{fig:mcp} provides a generic framework of FinAgents, powered by FinGPT and agent protocols.  It enables the development of multiple FinAgents tailored to various financial scenarios. We categorize the nine financial agents into two groups: tool agents and financial service agents.

\subsubsection{Tool Agents}
\begin{itemize}[leftmargin=*]
    \item \textbf{Search agent}: Facing massive multimodal financial data, the MFFM-enhanced FinGPT search agent retrieves and generates personalized results tailored to the diverse backgrounds and requirements of compound users. These agents would facilitate data-driven decision-making by providing precise, context-aware insights. More importantly, compared to professional financial database platforms such as the Bloomberg Terminal\footnote{https://www.bloomberg.com/professional/products/bloomberg-terminal/}, the cost of using commercial multimodal large language models (MM-LLMs) (e.g., GPT-4o) or deploying open-source MM-LLMs is lower. Users can easily construct their own customized financial AI search agents, achieving search results that rival those of professional agencies. The effectiveness of such an approach has already been demonstrated by FinGPT search agent~\cite{Felix2024FinGPTAgent, Zhang2023RetrievalAugmentedLLMs}.

    \item \textbf{Tutor agent}: There are two recent Guinness World Records \cite{guinness2024}: within a window of $24$ hours, $46,045$ attendance in an introduction to AI course and $112,718$ for a mathematics course, respectively. These numbers show a huge demand for online education. MFFMs can provide a scalable solution to meet this demand globally. For online education platforms, AI tutors equipped with the reasoning capabilities of MFFMs can provide high-quality tutoring services. For students, MFFMs can deliver a personalized learning experience. QFinben~\cite{Jaisal2024QFinBen} demonstrates that a pre-trained MFFM model with strong capabilities in undergraduate, graduate, and certification exams would provide a scalable, personalized solution for AI tutors in business and finance.

    \item \textbf{Robo-advisor}: Robo-advisors offer automated, algorithm-driven financial planning and investment management with minimal human intervention. They deliver personalized investment advice and portfolio management to individuals at a lower cost than traditional financial advisors. MFFMs can further enhance Robo-advisors by improving personalized interactions, integrating multimodal data for a comprehensive view of market and portfolio impacts, and providing ongoing adjustments and reminders through continuous user engagement.

    \item \textbf{Coding agent}: Coding agents empower investors to rapidly build personal financial analytical tools~\cite{wermelinger2023using}. 
    
\end{itemize}

\subsubsection{Financial Service Agents}

The financial services industry relies on digitized financial information for critical business decisions, such as business operations, investment, and mergers and acquisitions. Digitized financial information includes text, audio, images, and diverse market information. MFFM-powered workflow can integrate diverse multimodal financial data and offer customized financial services tailored to specific needs.

\begin{itemize}[leftmargin=*]
    \item \textbf{Credit scoring agent}: Leveraging LLMs, investors could build a credit scoring agent to generate transparent, data-driven credit scores. 

    \item \textbf{Auditing agent}: In auditing, auditors need to review lots of documents, manage multiple subtasks, and ultimately complete the auditing process. AI agents can autonomously perform complex audit procedures involving tasks such as AI-driven risk assessment or financial statement reviews~\cite{schreyer2024artificial}. By assisting auditors in completing these tasks, AI agents can improve auditing efficiency and reduce human error.

    \item \textbf{Compliance agent}: Integrating MFFMs into AI compliance offers organizations a scalable approach to managing regulatory and ethical requirements. Such an integration streamlines compliance workflows, automates complex regulatory analyses, and reinforces ethical AI practices—essential steps for building trust, mitigating risks, and fostering responsible AI developments~\cite{Keyi2024Regulations}.

    \item \textbf{Report generation agent}: Report generation refers to the use of MFFM-powered AI agents to consolidate complex financial data into concise, readable textual content. Regular, accurate, and insightful reports help stakeholders understand performance trends, identify risks, and make informed decisions. MFFM-powered report generation agents enable users to quickly generate high-quality, data-driven, and personalized financial reports.

    \item \textbf{Trading agent}: Trading is a complex financial decision-making task influenced by a wide array of market data. MFFM-powered agents can integrate different multimodal market information and output appropriate trading strategies. More importantly, it enables market stakeholders to employ agent systems to get personalized investment suggestions at a low cost. The Financial reinforcement learning (FinRL) trading agent \cite{liu2018practical, liu2020finrl, liu2021finrl, liu2022finrl} offers a user-friendly virtual market environment that includes a wide range of multimodal market data. FinRL integrates commonly used Deep Reinforcement Learning (DRL)  algorithms such as DQN, DDPG, PPO, SAC, A2C, and TD, enabling users to develop their own DRL trading strategies. Recently, FinRL 2025 contest\footnote{\url{https://finrl-contest.readthedocs.io/en/latest/}} proposed the FinRL-DeepSeek project, combining reinforcement learning with LLMs to develop an automated stock trading agent trained on stock price and financial news data. This hybrid approach enhances the capacity to process complex, evolving market information~\cite{wang2023alpha}.
    
    There are two LLM-based trading agent demos: 1) FinMem~\cite{yu2024finmem} is a single-agent system combined with a memory database to retain valid market information and adjust trading strategies in a timely manner. 2) FinCon~\cite{yu2024fincon} is a multi-agent framework that can handle multimodal market data, including text, time series, and audio. By employing a manager-analyst hierarchy, FinCon enables coordinated, natural language interactions and enhances decision-making with a unique self-critiquing mechanism for systematic investment belief updates.

\end{itemize}

By summarizing the application scenarios outlined above, it becomes clear that the MFFM-powered agent has great potential to offer market stakeholders scalable, personalized, and cost-effective solutions to multiple complex real-world financial tasks.





\begin{table*}[h!]
\centering
\begin{tabular}{l|c|c}
\toprule
\textbf{Characteristic} & \textbf{BloombergGPT} \cite{Wu2023BloombergGPT} & \textbf{FinGPT} \cite{Liu2023FinGPT, Liu2024FinGPTHPC, Yang2023FinGPT,zhang2023instruct} \\
\midrule
Backbone Model & Bloom & Llama \\
Model Parameters & 50 Billion &  8 Billion \\
Corpus Size (Tokens) & 708 billion (363B financial, 345B general) & Real-time fetching from 34 sources \\
Compute Resources & 512 NVIDIA A100 GPUs & Single NVIDIA RTX 3090 GPU \\
Total GPU-hours & $\sim$650,000 hours & $\sim$ 4-6 hours \\
Estimated Cost & \$2.76 million & Minimal (<$\$100$) \\
Update Frequency & One time  & Frequent fine-tuning \\
\bottomrule
\end{tabular}
\caption{Comparison of BloombergGPT and FinGPT in terms of training data and computation costs.}
\label{tab:fin_vs_bloom}
\vspace{-0.2in}
\end{table*}

\subsection{Enablers for a Coming Era of Agentic AI Ecosystem}

\subsubsection{Open Models.}  Open models encourage people's choice and utilization in building agentic AI systems. The openness of a model can be assessed from three dimensions: code, data, and documentation. Many models only open-source a portion of them, and the behavior of misusing the “open source” label is called openwashing~\cite{heimstadt2017openwashing, widder2024open, liesenfeld2024rethinking}. Openwashing poses a challenge as it introduces confusion into the agentic AI ecosystem.  

Model Openness Framework \cite{white2024model}\footnote{https://isitopen.ai/} and OpenMDW License~\footnote{https://openmdw.ai/} provide the framework to develop an open-source agreement, build an open-source standard, and evaluate the openness of models using a leaderboard~\cite{white2024model}. This approach would enhance transparency in model usage while preserving the integrity of the entire agentic AI ecosystem.

\subsubsection{Agent Protocols} Agent protocols specify message structures, negotiation mechanisms, and coordination procedures to facilitate efficient collaboration among agents. Currently, there are two commonly used protocols:
\begin{itemize}[leftmargin=*]
\item \textbf{Model Context Protocol (MCP)\footnote{https://www.anthropic.com/news/model-context-protocol}}: MCP is an open standard that enables developers to build secure, two-way connections between data sources and AI-powered tools.
\item \textbf{Agent to Agent (A2A)\footnote{https://developers.googleblog.com/en/a2a-a-new-era-of-agent-interoperability/}}: The A2A protocol will allow AI agents to communicate with each other, securely exchange information, and coordinate actions on top of various enterprise platforms or applications.
\end{itemize}

\subsection{Case Study of FinGPT-Powered Agents}
\subsubsection{FinGPT Search Agents}
FinGPT search agent \cite{Felix2024FinGPTAgent} can quickly retrieve multimodal financial data customized to the specific needs of individual users or institutional investors and generate personalized content. Fig. \ref{fig:mcp} provides a generic framework of the FinGPT-powered agents. Interaction begins through a user interface where users input inquiries. Then, the agent will call different MCP clients accordingly to communicate via standard input/output with the corresponding MCP servers. 

These MCP servers handle different functions such as: 1) 
accessing local financial files; 2) searching financial news from remote services like Yahoo Finance through Web APIs; 3) querying market data from databases; 4) analyzing the market sentiment from social platforms like Twitter and Reddit/WallStreetBets. The framework emphasizes user-controlled permissions, explicitly asking for authorization before accessing external data sources, thus maintaining transparency and user privacy. MCP includes local and remote interactions, with remote servers interacting through the networking protocol (HTTP with SSE), ensuring real-time data updates.


\begin{figure*}[ht]
\begin{center}
\centerline{\includegraphics[width=0.99\textwidth]{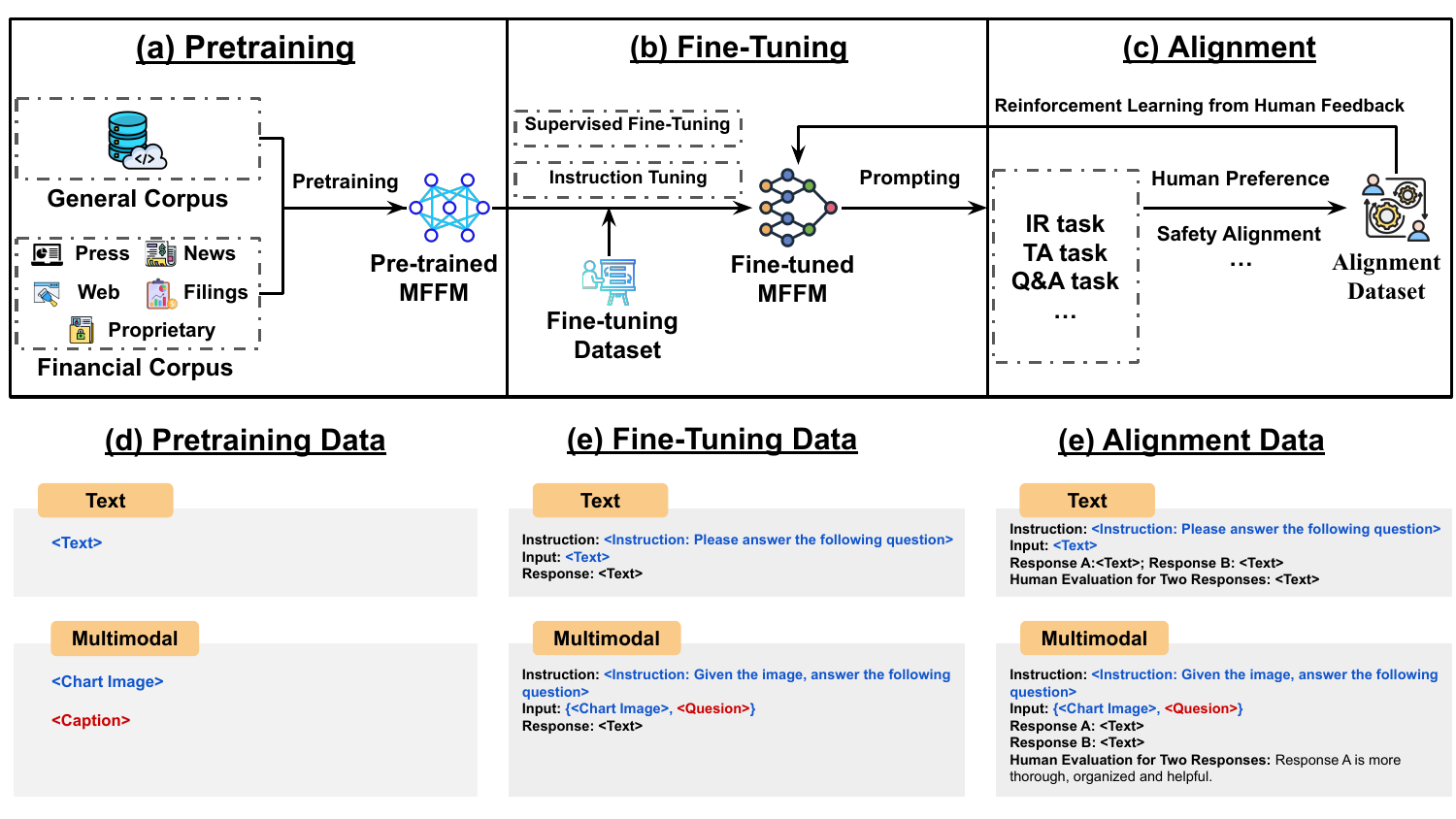}}
\caption{The reference lifecycle of model development. It consists of three key stages: (a) Pretraining, where a combination of general and financial corpora is used to pre-train the MFFM, ensuring that it comprehensively understands world financial knowledge; (b) Fine-Tuning, where the pre-trained MFFM fine-tuning on instruction or specific task dataset to enhance its understanding of user intentions or specific tasks; (c) Alignment Tuning, allowing the MFFM to generate content that is more human preference and secure. To make it easier to understand, we have provided examples of data from each stage in (d)–(e).}
\label{fig:lifecycle}
\end{center}
\vspace{-1em}
\end{figure*}

\subsubsection{Buffett Agent}
It is a fine-tuned FinGPT model that acts as a financial advisor in the style of Warren Buffett. Using the FinLoRA framework \cite{wang2024finlora}, such a FinGPT model was adapted by feeding in the Llama-3.1-8B base model with a custom `Buffett Brain' dataset of over 25,000 question–answer pairs. We curated this extensive dataset from diverse sources related to Buffett and Berkshire Hathaway, including Buffett’s Wikipedia pages and notable books about his life and investment philosophy. It also incorporates Berkshire Hathaway’s annual shareholder letters (to capture Buffett’s tone and advice), transcripts of the shareholder meetings, and even Berkshire’s SEC filings, ensuring the agent has both Buffett’s folksy tone and factual financial knowledge for technical questions.

The actual compute cost of the fine-tuning process was under \$20, indicating a highly accessible way to build a personalized Buffett agent. In the evaluations, the Buffett Agent’s responses aligned much more closely with Buffett’s style and verbiage than those of the base model, achieving roughly an 8.8\% higher BERTScore. The agent effectively produces plain-spoken, conversational advice reminiscent of Buffett’s own folksy wisdom. We provide a live demonstration of Buffett Agent at \footnote{\url{https://finlora-docs.readthedocs.io/en/latest/tutorials/buffett_agent.html}}. 

To further enhance the agent’s capabilities, Buffett Agent can integrate with FinGPT’s multi-source search architecture. By using FinGPT Search Agent MCP server, Buffett Agent is able to retrieve up-to-date Berkshire Hathaway letters, news headlines, and filings on demand. With such an upgrade, the Buffett Agent could maintain Buffett’s authentic voice while supplementing its answers with current information and deeper domain vocabulary.

\section{Landscape of MFFMs: Progress and Prospect}

Recent progress in foundation models with enhanced multimodal capabilities has attracted research efforts to explore their financial counterparts. First, through a case study, we present our own first-hand experiences of training a MFFM model and illustrate a reference lifecycle. Then, we summarize the latest development in the general-purpose domain. Subsequently, we describe the progress of MFFMs from three aspects: benchmark, model, and dataset. At the end, we highlight the prospects for MFFMs.


\subsection{First-hand Experiences of Model Training through Case Studies}
\label{ssec:case_study}

We first review the training experiences of two representative language-centric financial foundation models: BloombergGPT \cite{Wu2023BloombergGPT} and FinGPT \cite{Liu2023FinGPT, Liu2024FinGPTHPC, Yang2023FinGPT,zhang2023instruct}. Then, we  summarize our experiences of training Open-FinLLMs \cite{Xie2024OpenFinLLMs} and present a reference  lifecycle of MFFM development.

\subsubsection{Case Study of Language-centric Financial Foundation Model}

A high-quality financial corpus and a powerful base model are two key ingredients for the successful pretraining of a domain-specific foundation model. As a representative FinLLM, BloombergGPT \cite{Wu2023BloombergGPT} exemplifies large-scale pretraining in finance: it is a 50-billion-parameter model trained on a mixed dataset with 708 billion tokens of text (about 363B tokens from the financial domain and 345B tokens from the general domain). This mixed dataset allows the model to retain broad general knowledge while acquiring deep financial expertise. BloombergGPT outperforms general-purpose LLMs on financial tasks without sacrificing performance on general tasks.

Pretraining requires prohibitive computational resources and training data. In contrast, FinGPT \cite{Liu2023FinGPT} took a data-centric approach with LoRA fine-tuning. It aggregates real-time financial text from 34 diverse Internet sources and uses parameter-efficient tuning methods (e.g., LoRA, QLoRA) to adapt a general LLM to financial tasks. By leveraging continuously updated data and even incorporating market feedback through a reinforcement learning strategy based on stock price movements, FinGPT can continually refine a model’s financial knowledge at relatively low computational cost. This approach adds flexibility and keeps the model’s expertise up-to-date, complementing the one-time fine-tuning of specific datasets.

\textbf{FinGPT vs. BloombergGPT: Data and Cost Comparison}. Table \ref{tab:fin_vs_bloom} provides a comparison between BloombergGPT and FinGPT. BloombergGPT utilized a massive corpus exceeding 700 billion tokens and required approximately 650,000 GPU-hours, resulting in an estimated cost of \$2.7 million. In contrast, FinGPT employs a fine-tuning approach on updated financial datasets, reducing both the data requirements and computational expense. For the sentiment analysis task on the Financial Phrase Bank dataset, FinGPT's F1-score of 0.878 beats BloombergGPT's 0.511. FinGPT offers a cost-efficient solution and enables frequent updates of the model for diverse financial applications.

\subsubsection{Case Study of Training a MFFM Model}

The development of MFFMs may consist of three stages: pretraining, fine-tuning, and alignment, where a reference lifecycle is shown in Fig. \ref{fig:lifecycle}.  Based on our experiences with the project Open-FinLLMs \cite{Xie2024OpenFinLLMs}, we elaborate on each stage accordingly.

\textbf{(Continual) Pre-training stage}: Open-FinLLMs \cite{Xie2024OpenFinLLMs} employ an 18 billion-token corpus from the general domain and a 52 billion-token corpus from the financial domain. This curated dataset allows the model to keep the general knowledge while getting the financial knowledge. Then, Open-FinLLMs chose Llama3-8B as the base model for the continual pre-training and obtained a financial model called FinLLaMA. The continual pre-training process runs on $64$ A100 80GB GPUs, approximately $250$ GPU hours per epoch. FinLLaMA set the maximum sequence length to $8,192$ tokens. FinLLaMA surpasses its base model LLaMA3-8B on several financial tasks, highlighting the effectiveness of (continual) pretraining.

\textbf{Fine-tuning stage}: This step aims to enable the model's multimodal capabilities, enhance the model’s instruction-following capabilities, and optimize performance on downstream financial tasks. Building upon FinLLaMA,  its multimodal extension, FinLLaVA addresses multimodal financial tasks by employing multimodal instruction tuning. The instruction-tuning dataset comprises $1.43$ million image-text pairs. Instruction tuning is conducted on eight NVIDIA HGX H20 80GB GPUs, with the entire process requiring approximately 30 hours for one epoch. FinLLaVA outperforms all open-source MM-LLMs chart understanding tasks and is second only to the closed-source GPT-family MM-LLMs.

\textbf{Alignment stage}: This step aims to guide fine-tuned MFFMs to generate human-preferred and safety output. FinTral~\cite{bhatia2024fintral} includes an alignment tuning process. 
First, an alignment dataset is constructed. FinTral fed the instruction dataset into both a high-capacity LLM, such as GPT-4, and a less capable casual LLM. The output from the high-capacity LLM is labeled as positive samples, while those from the casual LLM are labeled as negative samples. Then, alignment tuning is conducted using the Direct Preference Optimization (DPO) method~\cite{rafailov2023direct}. After alignment tuning, FinTral not only generates the output that is aligned with human preferences but also greatly reduces the hallucinatory content. 

\begin{table*}[t]
\resizebox{\textwidth}{!}{
\begin{tabular}{l|>{\columncolor{cyan}}c|ccccccc|cccccc|ccc}
\toprule
& & \multicolumn{7}{c|}{Text Tasks}                  &  \multicolumn{5}{c}{Multimodal Tasks} &                        & \multicolumn{3}{c}{Features} \\ \midrule
\multicolumn{1}{c|}{Benchmark} 
 & \textbf{\#Tasks}                
& IE & TA & QA & RM & FO & DM & CQA & VQA & CU & NU & IU & I2T & ASR  & RAG  & Agent  & Language \\ 
 \midrule
FinBen \cite{Xie2024FinBen}  & \multicolumn{1}{>{\columncolor{cyan}}c|}{\textbf{24}} & 6  & 8  & 3  & 4  & 1  & 1  & \multicolumn{1}{c|}{-} & -        & -      & -      & -   & -   & \multicolumn{1}{c|}{-} & -    & \textcolor{green}{\CheckmarkBold}    & EN           \\
MultiFinBen \cite{peng2025multifinben}     & \multicolumn{1}{>{\columncolor{cyan}}c|}{\textbf{28}} & 7  & 8  & 8  & 1  & 1  & 1  & \multicolumn{1}{c|}{-} & -        & -      & -      & -    & -  & \multicolumn{1}{c|}{3} & -    & \textcolor{green}{\CheckmarkBold}     & EN/ZH/ES/GR/Jpn        \\
QFinBen \cite{Jaisal2024QFinBen} & \multicolumn{1}{>{\columncolor{cyan}}c|}{\textbf{1}}  & -  & -  & -  & -  & -  & -  & \multicolumn{1}{c|}{1} & -        & -      & -      & -    & -   & \multicolumn{1}{c|}{-} & \textcolor{green}{\CheckmarkBold}  & -      & EN           \\
OmniEval  \cite{wang2024omnieval}                    & \multicolumn{1}{>{\columncolor{cyan}}c|}{\textbf{5}}  & 5  & -  & -  & -  & -  & -  & \multicolumn{1}{c|}{-} & -        & -      & -      & -    & -   & \multicolumn{1}{c|}{-} & \textcolor{green}{\CheckmarkBold}  & -      & EN           \\
InverstorBench  \cite{li2024investorbench}              & \multicolumn{1}{>{\columncolor{cyan}}c|}{\textbf{3}}  & -  & -  & -  & -  & -  & 3  & \multicolumn{1}{c|}{-} & -        & -      & -      & -  & -    & \multicolumn{1}{c|}{-} & -    & \textcolor{green}{\CheckmarkBold}    & EN 
\\
\midrule
FFAMA   \cite{xue2024famma}                      & \multicolumn{1}{>{\columncolor{cyan}}c|}{\textbf{1}}  & -  & -  & -  & -  & -  & -  & \multicolumn{1}{c|}{-} & 1        & -      & -      & -    & -  & \multicolumn{1}{c|}{-} & \textcolor{green}{\CheckmarkBold}  & -      & EN/ZH/FN \\
MME-Finance \cite{gan2024mme}                  & \multicolumn{1}{>{\columncolor{cyan}}c|}{\textbf{10}} & -  & -  & -  & -  & -  & -  & \multicolumn{1}{c|}{-} & 1        & -      & 3      & 3    & 3  & \multicolumn{1}{c|}{-} & -    & -      & EN           \\
FinSet-Benchmark  \cite{bhatia2024fintral}            & \multicolumn{1}{>{\columncolor{cyan}}c|}{\textbf{9}}  & 1  & 2  & 1  & 1  & 1  & -  & \multicolumn{1}{c|}{-} & 1        & 1      & 1      & -   & -   & \multicolumn{1}{c|}{-} &   -   &   -     & EN           \\
FinAudio  \cite{cao2025finaudio}     & \multicolumn{1}{>{\columncolor{cyan}}c|}{\textbf{3}} & -  & -  & -  & -  & -  & -  & \multicolumn{1}{c|}{-} & -        & -      & -      & -  & -    & \multicolumn{1}{c|}{3} & -    & -    & EN           \\
\midrule
\textbf{FinLLM Leaderboard}  \cite{Colin2024Leaderboard}     & \multicolumn{1}{>{\columncolor{cyan}}c|}{\textbf{24}} & 6  & 8  & 3  & 4  & 1  & 1  & \multicolumn{1}{c|}{-} & -        & -      & -      & -  & -    & \multicolumn{1}{c|}{-} & \textcolor{green}{\CheckmarkBold}    & \textcolor{green}{\CheckmarkBold}     & EN           \\

\bottomrule
\end{tabular}}
\caption{Comparison of financial benchmarks. Text tasks: Information Extraction (IE), Text Analysis (TA), Question Answer (QA), Risk Management (RM), Forecasting (FO), Decision-Making (DM), Complex Question Answer (CQA). Multimodal tasks: Visual Question Answer (VQA), Chart Understanding (CU), Numeral Understanding (NU), Image Understanding (IU), Image-to-Text (I2T), Automatic Speech Recognition (ASR).}
\vspace{-2em}
\end{table*}

\subsection{Progress of MFFMs}

Taking a model consumer perspective, we review the progress of MFFMs from three aspects: evaluation and benchmarking suite, model development, and multimodal financial datasets.

A generic workflow would be as follows. A model consumer starts by selecting a set of benchmark questions that captures the expected model capabilities. Then, she may evaluate a set of candidate models (closed models via APIs and open models via publicly accessible weights) on such benchmark questions and pick the best-performing model. Lastly, she may curate the multimodal financial dataset to fine-tune the selected model.

\noindent
\subsubsection{\textbf{Evaluation and Benchmarking Suite}} 
Measuring a model's performance on various financial tasks is crucial for understanding its quantitative capabilities. Benchmarks are used for model comparisons, evaluations are used to understand the performance properties of the system, and tests are used to validate that those properties fall within acceptable bounds. 

\textbf{Benchmarking Question Sets}. Currently, multiple financial benchmarks provide comparisons from different perspectives. 
\begin{itemize}[leftmargin=*]
    \item \textbf{FinBen} \cite{Xie2024FinBen}\footnote{The name ``FinBen" also implies the big bang moment of powerful FinLLMs and FinAgents.}: It includes $46$ datasets spanning $24$ financial tasks and covers seven critical tasks: information extraction (IE), textual analysis, question answering (QA), text generation (TG), risk management (RM), forecasting (FO), and decision-making (DM). FinBen evaluated 30 representative LLMs and identified several key findings: LLM performed well in IE, and text analysis, but its performance in complex tasks such as high-level reasoning and text generation and prediction still needs to be improved.

    \item \textbf{MultiFinBen} \cite{peng2025multifinben}: Extending FinBen, MultiFinBen is the first multilingual and multimodal benchmark over the global financial domain. It evaluates LLMs across multiple modalities (text, vision, and audio) and various linguistic settings (monolingual, bilingual, and multilingual) on domain-specific tasks. Extensive evaluations of 22 SOTA models show that despite their strong general capabilities, these models perform poorly on complex cross-lingual and multimodal financial tasks.
    
    \item \textbf{High-quality financial benchmark (QFinBen)~\cite{Jaisal2024QFinBen}:} QFinBen explores the reasoning capabilities of LLM in complex financial questions. QFinBen assembled a dataset of 8,050 questions sourced from undergraduate and graduate finance, accounting, and economics examinations alongside professional financial exams such as the CFA, CPA, and FRM. QFinBen tests the dataset by using the GPT-4o, Llamma 3.1-405B, and Mistral Large 2. The findings indicate that LLMs still struggle to pass these complex examinations, highlighting their current limitations in addressing sophisticated financial reasoning challenges.

    \item \textbf{OmniEval~\cite{wang2024omnieval}:} It is the first RAG benchmark in the financial domain. OmniEval evaluates the RAG framework from a multi-dimensional that includes 1) a matrix-based RAG evaluation system that classifies queries into five tasks and 16 financial topics, thereby structuring the evaluation of different query scenarios; 2) A multi-stage evaluation system that evaluates search and generation performance to evaluate the RAG process comprehensively; and 3) Robust evaluation metrics derived from rule-based and LLM-based evaluation metrics. The results of OmniEval highlight that the RAG can effectively integrate external knowledge to improve the accuracy of the generated results in a variety of tasks. However, the evaluations also reveal that the RAG system struggles with complex multi-hop reasoning and numerical understanding.

    \item \textbf{InverstorBench~\cite{li2024investorbench}:} It's the first LLM-based financial agent benchmark. InverstorBench provides a comprehensive performance evaluation of 13 different LLMs across varied market scenarios, including stock trading, cryptocurrency trading, and ETH trading. This benchmark shows that proprietary models (e.g. GPT-4) generally exhibit better financial decision-making capabilities under complex market conditions. However, InverstorBench also indicates that the performance of different LLMs varies in stock, cryptocurrency, and ETF trading. This variability not only underscores the inherent complexity of financial markets but also emphasizes the critical importance of model selection and fine-tuning on specific financial corpus. 

    \item \textbf{FFAMA \cite{xue2024famma}:} It's an open-source benchmark for financial multilingual multimodal question answering (QA). It includes 1,758 meticulously collected question-answer pairs from university textbooks and exams, spanning 8 major subfields in finance, including corporate finance, asset management, and financial engineering. FFAMA assesses a variety of SOTA MM-LLMs, revealing that FAMMA presents a considerable challenge for these MM-LLMs. Even advanced models such as GPT-4o and Claude35-Sonnet attain only a 42\% accuracy.
    
    \item \textbf{MME-Finance \cite{gan2024mme}:} MME-Finance is a bilingual financial visual question and answer (VQA) benchmark. MME-Finance conducted extensive experimental evaluations on 19 MM-LLMs to test their perception, reasoning, and cognitive abilities on financial multimodal data. The results show that MM-LLMs that perform well in general benchmark tests may perform poorly on MME-Finance. Specifically, these models show poor performance in understanding candle charts and technical indicator charts. 

    \item \textbf{FinSet-Benchmark~\cite{bhatia2024fintral}:} It's part of FinTral~\cite{bhatia2024fintral}, containing 13 LLMs on seven text-based financial tasks, and 9 MM-LLMs on Chart Understanding. FinSet-Benchamrk indicates that the post-trained model using reinforcement learning with AI feedback (RLAIF), like FinTral-DPO, ChatGPT, and GPT-4, shows significant enhancements in comprehending complex texts, identifying specific entities, and interpreting numerical data.
    
    \item \textbf{FinAudio \cite{cao2025finaudio}:} FinAudio is the first benchmark designed to evaluate the capacity of AudioLLMs \cite{wang2024audiobench} in the financial domain. It first defines three tasks based on the unique characteristics of the financial domain: 1) ASR for short financial audio, 2) ASR for long financial audio, and 3) summarization of long financial audio. Then, the authors curate two short and two long audio datasets, respectively, and develop a novel
    dataset for financial audio summarization. Then, FinAudio evaluates seven prevalent AudioLLMs. Our evaluation reveals the limitations of existing AudioLLMs in the financial domain and offers insights for improving AudioLLMs.

\end{itemize}

\begin{table*}[t]
\centering
\begin{tabular}{lccc|c}
\toprule
\textbf{Task Category} & \multicolumn{3}{c}{\textbf{Evaluation and Benchmarking Suite} } & \textbf{\#Questions} \\
\cmidrule(lr){2-4}
 & \textbf{FinBen} \cite{Xie2024FinBen} & \textbf{MultiFinBen} \cite{peng2025multifinben} & \textbf{Open FinLLM Leaderboard} \cite{Colin2024Leaderboard}  & \\
\midrule
 \multicolumn{5}{c}{\textit{\textbf{from classicial datasets}}}\\
Information Extraction  & 4.6k & 14k & from FinBen and MultiFinBen &  18.6k\\
Text Analysis  & 17k & 10k & from FinBen and MultiFinBen & 27k \\
Text Generation  & 3k & from FinBen & from FinBen & 3k \\
Question Answer  & 4.5k & 7.1k & from FinBen and MultiFinBen & 11.6k  \\
Forecasting & 4.8k & 1.7k & from FinBen and MultiFinBen & 6.5k \\
Risk Management & 77k & from FinBen & from FinBen & 77k \\
Decision-Making & 3.4k & 1k & from FinBen and MultiFinBen & 4.4k \\
\midrule
\multicolumn{5}{c}{\textit{\textbf{Newly created datasets}}}\\
SEC filing analysis & - & - & 35.9k & 35.9k \\
XBRL reporting & - & - & 24.2k & 24.2k \\
Financial regulations & - & - & 5.1k & 5.1k \\
Certification questions & - & - & 85.6k & 85.6k \\
\midrule
Total & 114.3k & 33.8k & \textbf{150.8k} & \textbf{300k} \\
\bottomrule
\end{tabular}
\caption{Questions by category with additional benchmarks.}
\label{tab:dataset_sizes}
\end{table*}

These benchmarks provide an overview of the current landscape of applying LLM to financial tasks. We can find that: 1) LLMs/MM-LLMs can effectively improve the capabilities of information extraction relevant tasks and basic financial text analysis. Such improvements can help users build automated financial data processing systems, thus saving manual efforts and reducing human errors; and 2) Current LLMs/MM-LLMs still have limitations in their capacity to answer complex financial questions, comprehend numerical values, and interpret charts and tables. This underscores the urgency of developing MFFMs tailored for financial multimodal data.

\noindent \textbf{Open FinLLM leaderboard} \cite{Colin2024Leaderboard} \footnote{\url{https://github.com/finos-labs/Open-Financial-LLMs-Leaderboard/}}: FinLLMs and FinAgents with multimodal capabilities are rapidly advancing, poised to revolutionize a wide range of applications across business, finance, accounting, auditing, etc. Therefore, the timely evaluation of newly developed FinLLMs and FinAgents is critical. Benchmarks are static and lack the momentum to continuously adapt to and evaluate emerging FinLLMs and FinAgents, thereby limiting their utility for real-world applications and innovations. Therefore, establishing a standardized, continuously maintained leaderboard is essential for the ongoing development and improvement of Multimodal FinLLMs and FinAgents. Building on PIXIU \cite{xie2023pixiu}, FinBen \cite{Xie2024FinBen} and MultiFinBen \cite{peng2025multifinben}, the Open FinLLM leaderboard aims to maintain a dynamically open platform that encourages innovative adoption and improved models. Open FinLLM Leaderboard provides an interface between academia, the open-source community, the financial industry, and other stakeholders. Open FinLLM Leaderboard creates a collaborative and open ecosystem by continuously updating new datasets, tasks, and model performance.

Table \ref{tab:dataset_sizes} summarizes the sizes of question sets currently available on the Open FinLLM leaderboard, categorized by financial tasks. The Leaderboard is designed not only to be compatible with existing benchmarks but also to dynamically incorporate new question sets. As shown in Table \ref{tab:dataset_sizes}, the Open FinLLM Leaderboard has recently introduced several new datasets comprising 150,800 novel questions. These additions include expert-level seniors such as SEC filing analysis, XBRL reporting, financial regulation comprehension, and certification-related queries. These additional questions reflect the practical requirements and specific contexts of the financial industry.

\begin{figure}[t]
\centering
\includegraphics[width=0.45\textwidth]{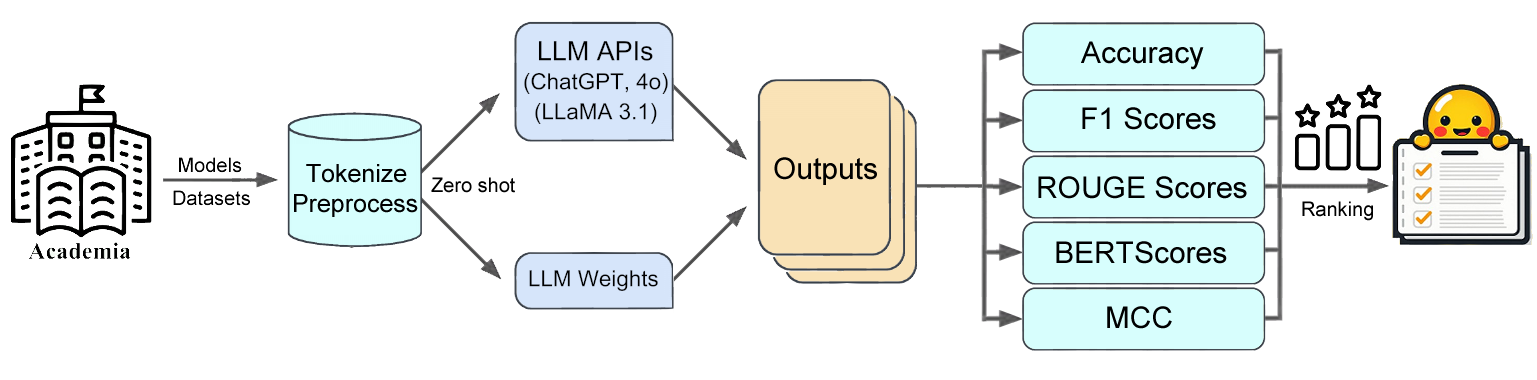}
\caption{Testing pipeline used in the FinLLM leaderboard \cite{Colin2024Leaderboard}.}
\label{fig:leaderboard}
\end{figure}

Such a broad coverage allows industry professionals to identify models suitable for specific applications, such as SEC filing analysis for market predictions or financial regulations for credit scoring and fraud detection. Fig. \ref{fig:leaderboard} is an overview of the testing pipeline. It employs a zero-shot evaluation setting to test expert-validated datasets, assessing models in multimodal settings across various financial tasks. Models are compared fairly based on their ability to handle unseen tasks in finance.

The Open FinLLM Leaderboard also aims to foster an open community that drives financial AI toward real-world applications and establishes a gateway between academia and industry. By translating complex research achievements into accessible and actionable insights, it aims to foster the growth of the Agentic AI Ecosystem. The Open FinLLM Leaderboard is similar to established industry standards such as MCP and MOF. It sets the benchmark for financial AI readiness, ensuring that innovations in financial language models are both
practical and impactful. The authors also discuss critical aspects and how this leaderboard and the surrounding community will contribute to FinLLMs’ readiness.

\begin{figure}[t]
    \centering
    \includegraphics[width=0.45\textwidth]{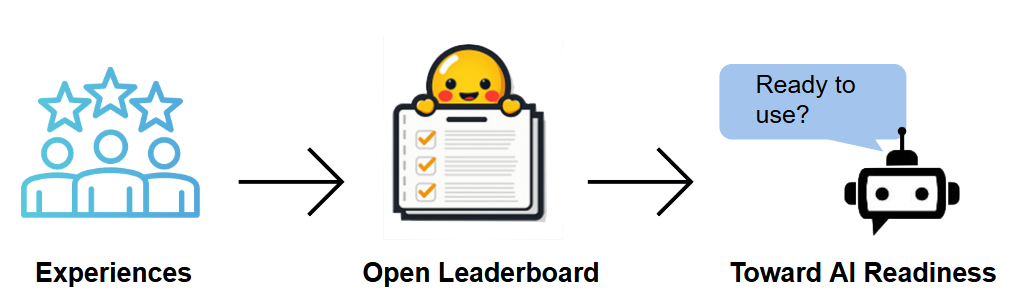}
    \caption{From user experiences to FinAI readiness: The iterative process of evaluating FinLLMs on the leaderboard \cite{Colin2024Leaderboard}.}
    \label{fig:Experience2Readiness}
\end{figure}

\noindent 
\subsubsection{\textbf{MFFM Models}}  Typically, MFFMs are built from open-source LLMs, which serve as the backbone. These MFFMs are pre-trained and fine-tuned using the specialized financial dataset. We aim to provide readers with a comprehensive understanding of the advancements in Multimodal Foundation Financial Models. We highlight representative MFFMs in below:
\begin{itemize}[leftmargin=*]
    \item \textbf{Open-FinLLMs \cite{Xie2024OpenFinLLMs}:} We have introduced Open-FinLLMs in Subsection \ref{ssec:case_study}. Open-FinLLMs consist of two Financial LLMs: FinLLaMA and FinLLaVA. The experiment results demonstrate that FinLLaMA gets superior performance over LLaMA3-8B, LLaMA3.1-8B, and BloombergGPT in text classification, credit scoring, fraud detection, Q\&A, Sentiment Analysis, NER, and decision-making tasks. FinLLaVA outperforms GPT4 and other Financial LLMs in understanding tables and charts. The results from Open-FinLLMs highlight the effectiveness of training financial domain-specific LLMs/MM-LLMs. 

    \item \textbf{FinTral \cite{bhatia2024fintral}:} It is a suite of state-of-the-art MFFMs built upon the Mistral-7B model and tailored for pure text and multimodal financial analysis. FinTral is pre-trained on 20 billion tokens of domain-specific data in the first step, followed by instruction fine-tuning and alignment with AI feedback. Subsequently, FinTral gets further instruction fine-tuning on multimodal instruction data. FinTral demonstrates good zero-shot capabilities, outperforming GPT-4 in five of eight text-based tasks. Moreover, FinTral's multimodal performance surpasses that of all other open-source MM-LLMs, ranking just behind GPT-4V.
    
    \item \textbf{FinVis-GPT~\cite{wang2023finvis}:} It's a new MFFM specialized in financial chart analysis. FinVis-GPT is pre-trained on a finance-oriented alignment/instructing following dataset, which includes various types of financial charts and their corresponding descriptions. The experiment results show that FinVis-GPT can interpret financial charts and provide valuable analysis. 
\end{itemize}

These studies have demonstrated that MFFMs already take important roles in multiple financial tasks. These works lay the foundation for more sophisticated applications of AI in finance, potentially transforming the landscape of financial analysis. Although the performance of these MFFMs in some complex financial tasks still needs to be improved, these findings also highlight the significant potential for future development of MFFMs. Future work will focus on further expanding the applicability of MFFMs in more complex tasks and diverse financial scenarios.

\noindent

\subsubsection{\textbf{Multimodal Financial Datasets}} Different stages rely on different types of training data. A high-quality training dataset will affect the capacity of the trained MFFMs. This part will discuss each stage's dataset construction and characteristics.
\begin{itemize}[leftmargin=*]
    \item \textbf{Pre-training dataset}. As the first stage, pre-training data aims to provide multimodal financial knowledge for models and enable the model to align the different modalities. During this stage, different models curate their unique training corpora. A representative training dataset is BloombergGPT's FinPile~\cite{Wu2023BloombergGPT}. It comprises a total of 345 billion tokens from public data and 363 billion tokens from proprietary data.
    
    \item \textbf{Instruction-tuning dataset}. This stage aims to teach models to better understand the instructions from the demanded tasks to boost zero-shot capacity. \textbf{OpenFinLLaVA} first assembled a comprehensive multimodal dataset, subsequently utilizing GPT-4o to extract financial content selectively. Ultimately, OpenFinLLaVA created an extensive collection of 662k multimodal pre-trained datasets, comprising images, charts, and tables. \textbf{FinVis-GPT} utilized historical daily data from Chinese A-share stocks to create visualizations, distributing the output into 80\% candlestick and 20\% line charts. \textbf{FinTral} utilizes several datasets to build a visual pretraining dataset. Additionally, the Llava Instruct dataset is employed to enhance instruction understanding in the multimodal LLMs, resulting in the creation of the instruction tuning dataset, FinVis-IT.
\end{itemize}


\subsection{Prospects of MFFMs} 


\subsubsection{\textbf{Reasoning Models.}} For financial reasoning tasks, foundation models must effectively understand domain-specific terminologies, analyze structured data (e.g., financial tables and charts), perform mathematical calculations, and extract insights from lengthy and complex documents \cite{xie2025finchain}. CFA questions also require strong reasoning capabilities \cite{nitarach2025fincot}. Those tasks place higher demands on the reasoning capabilities of foundation models. 

Reinforcement learning algorithms have demonstrated substantial potential in enhancing the reasoning capabilities of foundation models, bringing them closer to human-level reasoning skills \cite{shao2024deepseekmath}. Fin-R1 distills a financial reasoning dataset (Fin-R1-Data) that includes 60,091 complete chain-of-thought (CoT) reasoning paths from several financial datasets encompassing diverse financial scenarios \cite{liu2025fin}. Then,  Fin-R1 is established based on Qwen2.5-7B-Instruct,
using supervised fine-tuning and the Group Relative Policy Optimization algorithm (GRPO)
to enhance the model’s reasoning capability and standardize its output format. Results indicate that Fin-R1 achieves superior performance on financial question-answer tasks. Fin-O1 further evaluated the effectiveness of using different RL algorithms (DPO, PPO, GRPO) to improve model reasoning capabilities. The study found that GRPO consistently yields reliable gains, whereas PPO and DPO were less effective. These results highlight the importance of targeted data and specialized optimization over merely increasing model scale \cite{qian2025fino1}. 

These findings underscore the necessity for specialized data, tailored models, and domain-specific optimization techniques.


\subsubsection{\textbf{Multimodal retrieval-augment generation} (MRAG)} The ability to retrieve relevant information efficiently from a large database is crucial for the success of FinAI systems. Enhancing retrieval-augmented generation capabilities will enable more precise and contextually aware responses from AI models, significantly improving their usefulness in complex financial decision-making processes.

\subsubsection{\textbf{Customizing pretrained models to use scenarios}} Customizing pre-trained models to specific scenarios can significantly enhance user experience. For instance, trained on a curated dataset, the Buffett agent could serve as a robo-advisor in Buffett style \cite{wang2024finlora}. Another example is FinGPT search agent that provides personalized service to users for real-time information retrieval.


\subsubsection{\textbf{Fine-tuning and quantization methods}} For general-purpose LLMs to be effective in finance, they need to be fine-tuned with domain knowledge that captures the nuances of financial markets and instruments. Additionally, model quantization should be considered to optimize inference performance in terms of speed and resource consumption, ensuring that the models can be deployed effectively in real-time environments. FinGPT-HPC~\cite{Liu2024FinGPTHPC} and FinLoRA~\cite{ wang2024finlora} are two examples of applying quantization techniques in the fine-tuning process.

\subsubsection{\textbf{Mixture of Experts (MoE)}} The MoE architecture replaces a single large foundation model with multiple smaller specialized models \cite{shazeer2017outrageously, cai2024survey}. It decomposes complex problems into simpler tasks, with different models collaborating to determine the assignment of each input. This approach enables the model to handle larger input data volumes without increasing computational costs. Therefore, MoE enables greater model capacity and improved scalability, making it feasible to develop large multimodal financial foundation models with efficiency \cite{joshi2025survey}.

\subsubsection{\textbf{Federated Learning}} FL offers two benefits: 1) protecting the privacy of proprietary financial data \cite{byrd2020differentially, long2020federated}; and 2) conserving computing resources during the fine-tuning stage \cite{liu2023efficient}. In a federated learning environment, the DP-LoRA method \cite{liu2025differentially} provided a LoRA-based fine-tuning method for financial and medical scenarios while adding noise to protect data privacy.

\section{Challenges and Opportunities: Secure FinAI Readiness and Governance}

Adopting MFFMs in real-life scenarios will face several challenges, while presenting research opportunities. 

\subsection{Proprietary Multimodal Financial Data}

Proprietary data are important for financial analysis and decision-making because they provide unique insights.
\begin{itemize}[leftmargin=*]
    \item \textbf{Internal trading data}: The financial institutions have the capability to track and analyze their transaction data, offering insights into behavioral patterns and market trends.
    \item \textbf{Credit scoring data}: Financial entities possess data regarding the credit histories of individuals and corporations, which is essential for risk management.
    \item \textbf{Market research data}: Data gathered through specialized market research or customer feedback can aid financial firms in understanding consumer demands and market dynamics.
    \item \textbf{Real-time streaming data}: Certain institutions have access to real-time transaction flow data, which significantly facilitates algorithmic trading.
    \item \textbf{Private financial reports:} Some companies may have access to confidential financial information about partners or potential investment targets.
    \item \textbf{Proprietary economic indicators}: Large institutions may develop their own macroeconomic or microeconomic indicators based on exclusive datasets and analyses.
    \item \textbf{Alternative data:} This includes satellite imagery, mobile app data, and social media activities, which can provide additional perspectives and information for investment decisions.
\end{itemize}



\textbf{Synthetic Multimodal Data}. The training of MFFMs has two challenges: 1) Data privacy - the sensitivity of financial data limits its use in constructing training datasets; 2) Data quality - a scarcity of high-quality multimodal financial data, with the existing data mainly consisting of <Chart Image - Text> pairs that lack balanced representation from various modalities.  These challenges constrain the further development of MFFMs' capabilities. Therefore, enhancing the diversity and quality of multimodal financial data has become a critical need. Synthetic Multimodal Data provides a potential solution to these issues.

Synthetic data \cite{assefa2020generating} is from a generative process that learns the properties of real data but cannot be traced back to the raw data sources. The objective of synthesizing multimodal data is to generate data that accurately reflects the real distribution while also ensuring it cannot be traced back to the original sources to fulfill privacy requirements. 
There have been multiple demos in the medical field that have used synthetic multimodal data to augment datasets, which demonstrates the effectiveness of synthesizing multimodal data~\cite{wendland2022generation, pezoulas2024synthetic}. However, in the financial domain, \citet{potluru2023synthetic} provides a comprehensive review of the field of financial data synthesis and points out the current lack of efficient multimodal data synthesis methods. This highlights the challenges and opportunities of synthetic multimodal financial data.


\subsection{Digital Regulatory Reporting (DRR)}

A chatbot with multimodal capabilities \cite{Keyi2024Regulations}\cite{Felix2024FinGPTAgent} helps automate the financial regulatory process. For example, when lawyers perform case studies, chatbots can quickly search and summarize relevant legal provisions and historical cases, saving time over manual searches. When accountants prepare financial statements, chatbots can assist in checking compliance with generally accepted accounting principles (GAAP). 

However, the financial regulatory landscape presents unique challenges to MFFMs. First, the complex framework and overlapping jurisdictions of financial regulation make the compliance process complex. In the European Union (EU), the European Supervisory Authorities (ESAs) need to collaborate closely with national regulators to maintain a cohesive regulatory environment across Member States ~\cite{eu2022report, olha2021twosides}. The U.S. financial regulatory framework is fragmented, comprising federal and state laws. It involves various entities, including federal agencies, state regulators, interagency bodies, and international regulatory fora, with overlapping jurisdictions ~\cite{labonte2023framework}. Second, financial regulation requires processing multimodal data from different sources. This includes structured data, such as SQL databases and XBRL filings; unstructured data, such as regulatory texts; dynamic and noisy transaction data; and code in financial product management systems. The format and complexity of each data type vary greatly, which creates a challenging environment for AI compliance.

\textbf{XBRL}: eXtensible Business Reporting Language (XBRL) is an open international standard for business reporting, in order to streamline financial data creation, dissemination, and analysis. XBRL facilitates information exchange among investors, regulatory bodies, and market participants, boosting market transparency and regulatory compliance. Over the last two decades, most global economies have adopted XBRL for financial information sharing. However, the complexity of XBRL necessitates specialized knowledge for proper understanding and analysis, posing a steep learning curve for businesses and a challenge for widespread accessibility by the general public.

An XBRL agent \cite{Han2024XBRLAgent} will simplify data aggregation and support informed decision-making. It may provide users with easy access to financial intelligence. How we interact with financial data is no longer the exclusive domain of a few individual experts but a valuable resource for everyone. On the FinanceBench dataset, a public dataset comprising SEC document-related questions (150 openly available sample questions), \cite{Han2024XBRLAgent} evaluated the current AI ChatBots (e.g., ChatGPT, LLama2, FinGPT). The results show that its accuracy in answering financial questions is only about 19\% ~ 30\%, which is far from the professional level. The errors may come from several aspects: 1). Ambiguity in complex financial terminology; 2). Errors in interpreting and extracting data from financial documents; 3). Calculation errors (e.g., financial ratio calculation and aggregation).

\textbf{Common Domain Model (CDM)}: The Common Domain Model (CDM), a standardized, machine-readable/machine-executable data and process model for multiple financial products, is a promising fundamental solution to address the above challenges. Developing a CDM for XBRL using the Multimodal Large Language Model can handle various document formats, including PDFs, scanned documents, and webpages. High-quality document reading can effectively reduce errors during document reading and support financial documents in diverse scenarios. Furthermore, using MM-LLMs as the backbone and combining multiple external tools or RAG techniques to construct a standard agentic workflow can mitigate ambiguity in financial terminology and numerical calculation errors during format conversion.

\begin{figure*}[t]
\begin{center}
\centerline{\includegraphics[width=1.0\textwidth,  trim={0cm 2cm 0cm 2cm}, clip]{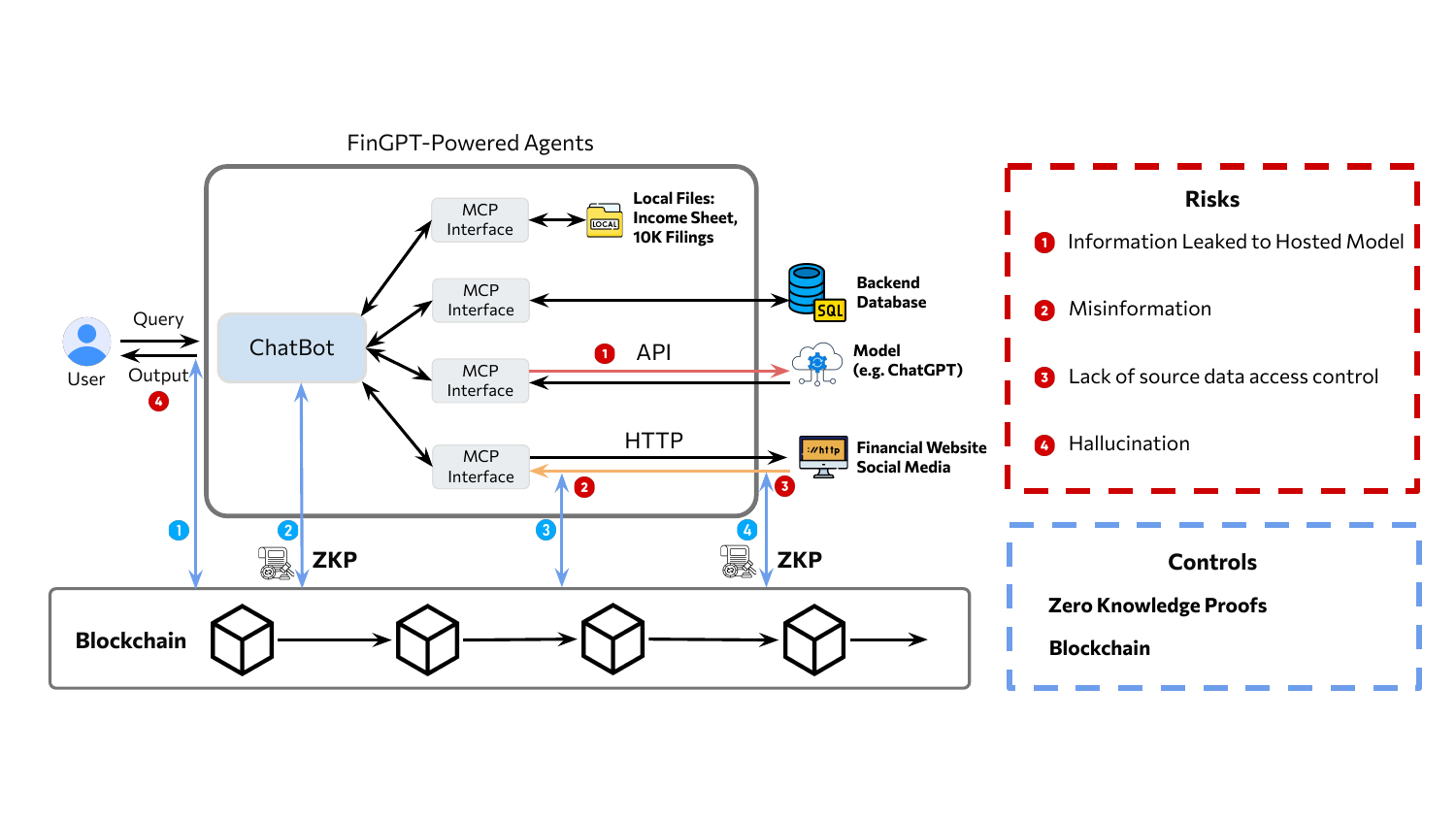}}
\caption{Guardrail framework for FinGPT-powered agents using Zero-Knowledge Proofs (ZKPs) and blockchain technologies.}
\label{fig:guardrail}
\end{center}
\vspace{-1em}
\end{figure*}




\subsection{Ethical Challenges}

There are intensified ethical concerns with MFFMs. Mishandling sensitive information and thus making unfair, biased judgments can be disastrous to financial institutions. Analysts who trust flawed MFFMs will make bad investment decisions and improper risk assessments. Small missteps can cause significant client dissatisfaction and negative media attention. 

\textbf{Persistent ethical issues} include: 
\begin{itemize}[leftmargin=*]
    \item Security and privacy: It is vital that FinLLMs have airtight security to prevent leakage of sensitive information. Example: Samsung employees accidentally leaked company secrets when prompting ChatGPT for help.
    \item Copyright infringement: FinLLMs trained on Internet data are not allowed to output copyrighted data to end users. Example: The New York Times sued OpenAI and Microsoft for using millions of its articles; Perplexity was accused of using articles from The Wall Street Journal or The New York Post to populate its RAG database and generate responses to user queries.
    \item Systematic bias: In decision-making processes, FinLLMs’ systematic bias may lead to unfair discrimination towards certain racial groups. According to Zillow and Consumer Reports, LLMs may quote African Americans at higher prices in home mortgages and car insurance due to historical segregation towards disaster-prone areas.
    \item Transparency, explainability, and accountability: It is important to ensure that FinLLMs are transparent, explainable, and accountable, providing clear responses, especially in finance where every decision has significant implications. J.P. Morgan Chase established its firmwide Explainable AI Center of Excellence (XAI COE) for research on explainability and fairness in finance.
\end{itemize}

\noindent \textbf{Newly-emerging ethical issues} include: 
\begin{itemize}[leftmargin=*]
    \item Truthfulness: LLMs consistently hallucinate, creating false statements. In business and finance, hallucinations are problematic because LLMs’ output must exactly match information extracted from earnings reports when queried. Microsoft faced backlash when Bing AI hallucinated when analyzing Gap and Lululemon’s earnings reports during a demo.
    \item Sycophancy: LLMs demonstrate sycophancy, catering their outputs to match user beliefs rather than being truthful. Sycophancy is problematic when it causes inaccurate confirmation of financial analysts' and accountants’ math.
    \item Compliance with professional norms: LLM responses must follow professional norms to avoid implicit toxicity in training data. This is vital to preserve company culture and public relations.
    \item Law and regulatory compliance: FinLLMs must comply with current financial laws and regulations when making decisions and chatting with end users. According to the Consumer Financial Protection Bureau, FinLLMs must comply with regulations in operations like fraud detection, citing concerns like discrimination against minority racial groups.
\end{itemize} 


\subsection{Misinformation and Hallucination} 

In the financial domain, the accuracy of information is important for the
integrity of market operations, risk management, compliance, and financial decisions. There are two sources of inaccurate financial information: dissemination of misinformation and hallucination from the model's output. 

Misinformation is from various media channels \cite{rangapur2023fin} and the misuse of LLMs to generate misinformation~\cite{zhou2023synthetic, cao2024can, cao2025cosmis}. Detecting financial misinformation is a challenge. To address this issue, FMDLlama \cite{liu2024fmdllama} fine-tuned the LLaMA-3 model on the Fin-Fact dataset \cite{rangapur2023fin} to detect financial misinformation. This case presents a feasible solution. By leveraging LLMs, an agentic framework can be developed to detect dynamically evolving financial misinformation.

Hallucination is factually incorrect output from LLMs due to their training on vast and diverse datasets. Ensuring the accuracy and reliability of LLM-generated outputs is crucial for their application in the financial industry. \citet{kang2023deficiency} quantified financial hallucinations and explored several potential solutions to mitigate them, including few-shot learning, decoding by contrasting layers, and RAG.

\subsection{Guardrail Framework for FinAgents}

Wide adoption of FinAgents raises concerns about privacy, security, and trust. We first outline the potential risks inherent in the FinAgent workflow. Then, we propose a guardrail framework that leverages zero-knowledge proofs (ZKPs) and blockchain technology. Blockchain and ZKPs ensure that FinAgents’ actions remain secure, verifiable, and immutable, fostering transparency and trust.

\subsubsection{\textbf{Threats and Risks in the FinAgent workflow.}} The three FinAgent prototypes in Section~\ref{ssec:finagent} follow an agentic pipeline: An agent calls various tools via the
MCP protocol to retrieve relevant content from local files, backend databases, remote models, and the Internet. We identify several major risks across different financial scenarios (\textcolor{red}{red points} in Fig. \ref{fig:guardrail}), referring to \textbf{Linux's \href{https://air-governance-framework.finos.org/}{AI Readiness Governance Framework}}:
\begin{itemize}[leftmargin=*]
    
    \item \textbf{Information leaked to host model}. Enterprise users may frequently employ FinGPT search agent to process local files (e.g., income sheet) for compliance tasks like internal audits, risk assessment, or regulatory reporting. These files contain personally identifiable information or commercially sensitive data. Corporate employees and students increasingly rely on AI tutors for financial knowledge from local textbooks. These books usually have copyright restrictions. During multi-round dialogues, the sensitive local files or copyrighted content may leak to external models (\textcolor{red}{red line} and point \textcolor{red}{\textcircled{1}} in Fig. \ref{fig:guardrail}).
    
    \item \textbf{Misinformation}. Users employ FinGPT search agent to obtain real-time information from financial websites and social media. However, the generated responses may contain misinformation and biased content (\textcolor{orange}{orange line} and point \textcolor{red}{\textcircled{2}} in Fig. \ref{fig:guardrail}).
    
    \item \textbf{Lack of source data access control}. FinGPT-powered agent could access external data sources (e.g., subscription-based websites). However, since these sources may enforce different access control policies, users might inadvertently access data that they are not authorized to retrieve directly from the original source. This unauthorized access may also lead to copyright issues (\textcolor{orange}{orange line} and point \textcolor{red}{\textcircled{3}} in Fig. \ref{fig:guardrail}). 
    
    \item \textbf{Hallucination}. LLM-based output may contain hallucination content, which refers to information that appears plausible but is factually incorrect. Inaccurate output can lead to costly errors, operational inefficiencies, and misinformed decisions (point \textcolor{red}{\textcircled{4}} in Fig. \ref{fig:guardrail}).
\end{itemize}

To mitigate these identified risks, integrating Zero-Knowledge Proof (ZKP) protocols and blockchain technologies represents a promising solution. These technologies form the foundation of a novel guardrail framework designed specifically for financial agents, as illustrated in Fig. \ref{fig:guardrail} (blue line, points \textcolor{blue}{\textcircled{1}} - \textcolor{blue}{\textcircled{4}}).

\subsubsection{\textbf{Zero Knowledge Proofs (ZKPs) for Privacy-Preserving}} 
Zero-Knowledge Proofs (ZKPs) are cryptographic protocols that let a \emph{prover} convince a \emph{verifier} of a statement’s correctness without disclosing any underlying secrets. The ZKPs ensure three key properties: \textbf{completeness} (an honest execution always produces a verifiable proof), \textbf{soundness} (no one can forge a valid proof without performing the correct computation), and \textbf{zero-knowledge} (the verifier learns nothing beyond the truth of the statement). zkLLM~\cite{sun2024zkllm} has demonstrated that ZKP protocols help protect the privacy of the large language model parameters (usually considered as intellectual property of model producers). For LLMs with 13B parameters, zkLLM can verify the inference process in less than 15 minutes, and the generated proof file has less than 200 KB.

To ensure privacy in the agent workflow, the agent generates a ZKP proof file and uploads it to the blockchain (\textcolor{blue}{blue line} \textcolor{blue}{\textcircled{2}}), demonstrating that the actions (search steps, inference steps, and output procedure) strictly adhere to pre-established inference schemes without exposing sensitive or proprietary data to remote model (\textcolor{red}{red line} and point \textcolor{red}{\textcircled{1}}). To enhance access control for external source data, the external participants generate a ZKP file for copyright (\textcolor{blue}{blue line} \textcolor{blue}{\textcircled{4}}). When the local agent takes actions, copyright permissions are granted by the blockchain, preventing unauthorized retrieval of external content (\textcolor{orange}{orange line} and point \textcolor{red}{\textcircled{3}}). 

\subsubsection{\textbf{Blockchain-Layered Agent Life Cycle.}} The generated ZKP protocol files, agent updates, copyright policies, and regulatory documents are recorded on a permissioned blockchain (e.g., Hyperledger Fabric or Corda). Participants interact with the blockchain by submitting cryptographic hashes referencing agent updates, inference steps, or compliance logs (\textcolor{blue}{blue line} \textcolor{blue}{\textcircled{1}} - \textcolor{blue}{\textcircled{4}}). Agents update trusted source lists (\textcolor{blue}{blue line} \textcolor{blue}{\textcircled{3}}) to avoid misinformation from external content (\textcolor{orange}{orange line}, point \textcolor{red}{\textcircled{3}}) and load inference schemes (\textcolor{blue}{blue line} \textcolor{blue}{\textcircled{1}}) to prevent hallucinations (point \textcolor{red}{\textcircled{4}}). This creates an immutable audit log, enabling stakeholders to verify agent compliance with approved procedures. 

These two components jointly enable FinGPT-powered agents to incorporate local data securely and produce on-chain verifiable proofs of correctness and compliance. By preserving confidentiality of sensitive data (via ZKPs) while anchoring essential references in a tamper-proof ledger (via blockchain), our approach harmonizes the conflicting needs of safety, confidentiality, transparency, and regulatory oversight.

\section{Discussion and Conclusion}

This paper offers a comprehensive overview of Multimodal Financial Foundation Models (MFFMs), highlighting their state of readiness. First, we review the multimodal financial data and application scenarios. Then, we describe the progress and future prospects of MFFMs. We further analyze the challenges and opportunities faced by MFFMs to achieve AI readiness.

By summarizing the current state of readiness, multimodal financial application scenarios, multimodal financial data, and the development of MFFMs, this paper aims to inspire future research and innovation in both the academic and financial industries.

As we navigate the integration of machine learning in business and finance, it is paramount to address the multifaceted challenges that arise from the unique characteristics of multimodal financial data and the new capabilities of MFFMs. Here, we outline strategic directions and considerations that will enhance the financial AI readiness for individuals and institutions:
\begin{itemize}[leftmargin=*]
    \item \textbf{Multilingual and multimodal}. Financial data is inherently complex, often presented in various modes, including text, numerical data, images, and more. An effective financial AI framework must be capable of interpreting and integrating these diverse multimodal data seamlessly. Furthermore, the global nature of finance demands multilingual capabilities to ensure that insights can be gleaned from data across different languages and regions. AI models should be equipped to handle multiple tasks simultaneously, such as risk assessment, fraud detection, and customer service, to provide comprehensive solutions. 
    \item \textbf{Open datasets and question sets}. Open datasets will facilitate the training of the more Powerful MFFMs. Adding complex open financial questions into training datasets can further enhance their reasoning capabilities. Furthermore, Public open datasets and question sets assist in establishing a standard benchmark for evaluating MFFMs.
    
    \item \textbf{Open leaderboard of MFFMs and FinAgents}. Building an open leaderboard enables rapid evaluation of the progress and characteristics of different MFFMs. It will facilitate the development of an agentic AI ecosystem.
    \item \textbf{Blockchain}. Data privacy and protection of model intellectual property are the challenges when developing an agentic AI ecosystem. Blockchain technology allows multiple organizations to collaboratively train a shared model while safeguarding data privacy, preventing leakage of model parameters, and transparently verifying each participant's contributions.
\end{itemize}

\begin{acks}
  The authors thank Keyi Wang for helping draft Fig. \ref{fig:mcp}. 

 Xiao-Yang Liu Yanglet acknowledges the support from Columbia's SIRS and STAR Program, The Tang Family Fund for Research Innovations in FinTech, Engineering, and Business Operations. Xiao-Yang Liu Yanglet also acknowledges the support from the NSF IUCRC CRAFT Center research grant (CRAFT Grant 22017) for this research. The opinions expressed in this publication do not necessarily represent the views of NSF IUCRC CRAFT.

\end{acks}


\bibliographystyle{ACM-Reference-Format}
\bibliography{ref}

\appendix

\input{appendix}

\end{document}

%% file: appendix.tex
\section{Terminology}

Multimodal Financial Foundation Models (MFFMs) is an intersection field of foundation models and finance. To facilitate readers from various backgrounds, Table~\ref{table:llm_terms} lists terminologies for foundation models and agents, while Table~\ref{table:finance_terms} for finance.

\begin{table*}[htb]
\centering
\small
\begin{tabular}{|l|p{13cm}|}
   \hline 
    \textbf{Key Terms} & \textbf{Explanations} \\
    \hline
    Transformer
    & A transformer is a neural network architecture that utilizes the multi-head attention mechanism.\\
    \hline
    Large Language Model (LLM)
    & LLM is a type of machine learning model for human-like text understanding and generation.\\
    \hline
    Pre-training
    & Pre-training refers to the initial training phase where a model learns general features from a large dataset.\\
    \hline
    \multirow{2}{*}{Fine-tuning}
    & Since a pre-trained LLM has a large number of parameters, trained on a huge dataset over millions of GPU hours, it is natural to employ a fine-tuning method to scale such a GPT model to hundreds of use scenarios. \\
     \hline
    Generative Pre-trained Transformer (GPT)
    & GPT is a family of LLMs based on a transformer architecture. \\
    \hline
    Prompt engineering & The process of structuring an instruction in order to produce the best possible output from an LLM model.\\
    \hline
    \multirow{2}{*}{Zero-Shot Prompting} & An LLM is given a task without examples or training on that task, relying on LLM's pre-existing knowledge to generate a response.  \\
    \hline
    Few-Shot Prompting & The prompt of providing a generative model with a few examples of a task to guide its output. \\
    \hline
    \multirow{2}{*}{Chain-of-Thoughts (CoT)} & A prompt engineering strategy to guide language models to handle complex reasoning tasks. For example, write the reasoning guidance in the prompt. \\
    \hline
    \multirow{2}{*}{In-Context Learning (ICL)}
    & ICL is a new learning paradigm where a language model observes a few examples and directly outputs the test input's prediction.\\
    \hline
    \multirow{2}{*}{Foundation Model}
    & A foundation model is a machine learning or deep learning model that is trained on vast datasets so it can be applied across a wide range of downstream tasks.  \\
    \hline
    FinLLM
    & A foundation model for financial applications.\\
    \hline
    \multirow{2}{*}{Multimodal} & Multimodal means ``having several modalities", and a "modality" refers to a type of input or output, such as video, image, audio, text, proprioception, etc.\\
    \hline
    \multirow{2}{*}{Retrieval-Augmented Generation (RAG)} & RAG is a process of optimizing the output of an LLM. It references an authoritative knowledge base outside of its training data sources before generating a response. \\ 
    \hline
    Low-Rank Adaptation (LoRA) & LoRA is a popular and efficient training technique that significantly reduces the number of trainable parameters.  \\ 
    \hline
    \multirow{2}{*}{QLoRA} & QLoRA is the extended version of LoRA, which works by quantizing the precision of the weight parameters in the pre-trained LLM to 4-bit precision. \\ 
    \hline
    \multirow{2}{*}{Agent} & A decision maker. The LLM-powered agent is a powerful framework for solving complex tasks by using an LLM as its central computational engine.\\
    \hline
    \multirow{2}{*}{Model Context Protocol (MCP)} & A standard for connecting AI assistants to the systems where data lives, including content repositories, business tools, and development environments. \\
     \hline
    Agent2Agent (A2A) Protocol & A standard to facilitate communication between independent AI agents.
 \\
        \hline
    Openwashing & A term used to describe the act of presenting a model as open source when it is not using a permissive license. \\ 
    \hline
    
\end{tabular}
\caption{Terminology for LLMs and agents.}
\label{table:llm_terms}
\vspace{-0.2in}
\end{table*}

\begin{table*}[htb]
\centering
\small
\begin{tabular}{|l|p{12.4cm}|}
   \hline 
    \textbf{Key Terms} & \textbf{Explanations} \\
    \hline
    Earnings Conference Calls (ECCs)
    &  A call between a public company and key stakeholders to discuss the company's financial results.\\
    \hline 
    Monetary Policy Calls (MPCs)
    & Countries' central banks hold MPC to decide what monetary policy action to take.\\
    \hline 
     \multirow{2}{*}{Environmental, Social, Governance (ESG)}
    & This is shorthand for an investing principle that prioritizes environmental issues, social issues, and corporate governance.\\
    \hline 
    Financial Decision Making
    & It encompasses evaluating options, making choices, and taking actions (trading) related to financial matters.\\
    \hline 
    eXtensible Business Reporting Language (XBRL)
    & XBRL is the global standard that powers digital reporting.\\
    \hline 
    \multirow{2}{*}{Common Domain Model (CDM)}
    & CDM is a standardized, machine-readable, and machine-executable data and process model for how financial products are traded and managed across the transaction lifecycle.\\
    \hline 
    \multirow{2}{*}{Robo-Advisor}
    & A type of automated financial advisor that provides algorithm-driven management services without human intervention.\\
    \hline 
    Digital Regulatory Reporting (DRR)
    & DRR is a cross-industry initiative to transform the reporting infrastructure.\\
    \hline 
    Greenwashing & Promotes false solutions to the climate crisis that distract from and delay concrete and credible action. \\
    \hline
\end{tabular}
\caption{Terminology for finance.}
\label{table:finance_terms}
\vspace{-0.2in}
\end{table*}